\pdfoutput=1

\documentclass[11pt]{article}

\usepackage[]{acl}

\usepackage{times}
\usepackage{latexsym}

\usepackage[T1]{fontenc}

\usepackage[utf8]{inputenc}

\usepackage{microtype}

\usepackage{inconsolata}

\usepackage{graphicx}
\usepackage{url}
\usepackage{float}
\usepackage{multirow}
\usepackage{color, colortbl}
\usepackage{arydshln}
\usepackage{array}
\usepackage{booktabs}
\usepackage{caption}
\usepackage{subcaption}
\usepackage{tikz}
\usepackage{amsmath}

\newcommand{\manner}{MANNeR}

\definecolor{Gray}{gray}{0.9}
\title{Train Once, Use Flexibly: A Modular Framework for \\ Multi-Aspect Neural News Recommendation}

\author{
 \textbf{Andreea Iana\textsuperscript{1}},
 \textbf{Goran Glavaš\textsuperscript{2}},
 \textbf{Heiko Paulheim\textsuperscript{1}},
\\
 \textsuperscript{1}Data and Web Science Group, University of Mannheim, Germany
\\
 \textsuperscript{2}CAIDAS, University of Würzburg, Germany
\\
 \small{
  \{andreea.iana, heiko.paulheim\}@uni-mannheim.de, goran.glavas@uni-wuerzburg.de
 }
}

\begin{document}
\maketitle
\begin{abstract}
Recent neural news recommenders (NNRs) extend content-based recommendation (1) by aligning additional \textit{aspects} (e.g., topic, sentiment) between candidate news and user history or (2) by diversifying recommendations w.r.t. these aspects. This customization is achieved by ``hardcoding`` additional constraints into the NNR's architecture and/or training objectives: any change in the desired recommendation behavior thus requires retraining the model with a modified objective. This impedes widespread adoption of multi-aspect news recommenders.  
In this work, we introduce \manner{}, a modular framework for \textit{multi-aspect} neural news recommendation that supports on-the-fly customization over individual aspects at inference time. With metric-based learning as its backbone, \manner{} learns aspect-specialized news encoders and then \textit{flexibly} and \textit{linearly} combines the resulting aspect-specific similarity scores into different ranking functions, alleviating the need for ranking function-specific retraining of the model. 
Extensive experimental results show that \manner{} consistently outperforms state-of-the-art NNRs on both standard content-based recommendation and single- and multi-aspect customization. 
Lastly, we validate that \manner{}'s aspect-customization module is robust to language and domain transfer.   
\end{abstract}

\section{Introduction}
\label{sec:introduction}

Neural content-based recommenders, trained to infer users' preferences from their click history, represent the state of the art in news recommendation \cite{li2019survey,wu2023personalized}. 
While previously consumed content clearly indicates users' preferences, \textit{aspects} other than content alone, namely categorical features of the news such as topical category, sentiment, news outlet, or stance, contribute to their news consumption decisions. 
Accordingly, some neural news recommenders (NNRs) leverage information on these aspects in addition to text content, be it (i) directly as model input \cite{wu2019naml, liu2020kred} or (ii) indirectly, as auxiliary training tasks \cite{wu2019tanr, wu2020sentirec}. 

Increased personalization is often at odds with \textit{diversity} \cite{pariser2011filter}. NNRs optimized to maximize congruity to users' preferences tend to produce suggestions highly similar in content to previously consumed news \cite{liu2021interaction, wu2020sentirec, sertkan2023effect}. Another strand of work thus focuses on increasing diversity of recommendations w.r.t. aspects other than content (e.g., sentiment). To this effect, prior work either (i) re-ranks content-based recommendations to decrease the aspectual similarity between them \cite{rao2013taxonomy, gharahighehi2023diversification}, or (ii) trains the NNR model by combining a content-based personalization objective with an aspect-based diversification objective \cite{wu2020sentirec, wu2022end, shi2022dcan, choi2022not}. 

Different users assign different importance to various news aspects (e.g., following developing events requires maximization of content-based overlap with the user's recent history; in another use-case, a user may prefer content-wise diversification of recommendations, but within the same topic of interest). Moreover, with personalization and diversification as mutually conflicting goals, users should be able to seamlessly define their own optimal trade-offs between the two. 
The existing body of work is ill-equipped for such multi-aspect customization, because each set of preferences -- i.e., to personalize or diversify for each aspect -- requires a different NNR model to be trained from scratch. Put differently, forcing global assumptions on personalization and diversification preferences (i.e., same for all users) into the model design and training prevents customization at inference time. 

\vspace{1.4mm}
\noindent\textbf{Contributions.} We propose a \textit{modular} framework for \textit{Multi-Aspect} Neural News Recommendation (\manner{}) to address this limitation. It leverages metric-based contrastive learning to induce a dedicated news encoder for each aspect, starting from a pretrained language model (PLM). This way, we obtain linearly-combinable aspect-specific similarity scores for pairs of news, allowing us to define ad-hoc at inference a custom ranking function for each user, reflecting their preferences across all aspects.
\manner{}'s modular design allows customization for any recommendation objective specified over (i) standard (i.e., content-based) personalization, (ii) aspect-based diversification, and (iii) aspect-based personalization. It also makes \manner{} easily extendable: to support personalization and diversification over a new aspect (e.g., news outlet), one only needs to train the aspect-specific news encoder for that aspect.    
Through extensive experiments
with \textit{topical categories} and \textit{sentiment} as additional aspects next to content itself, we find that \manner{} outperforms state-of-the-art NNRs on standard content-based recommendation.
Thanks to its module-specific outputs being \textit{linearly composable} between objectives, we show -- without training numerous models with different objectives -- that depending on the recommendation goals, one can either (i) vastly increase aspect diversity (e.g., over topics and sentiment) of recommendations or (ii) improve aspect-based personalization, while retaining much of the content-based personalization performance. Finally, we demonstrate that \manner{} with a multilingual PLM is robust to the (cross-lingual) transfer of aspect-based encoders.
\section{Related Work}
\label{sec:related_work}

\noindent\textbf{Personalized NNR.}
Neural content-based models have become the main vehicle of personalized news recommendation, replacing traditional recommenders relying on manual feature engineering \cite{wu2023personalized}. Most NNRs consist of a dedicated (i) news encoder (NE) and (ii) user encoder (UE) \cite{wu2023personalized}. The NE transforms input features into news embeddings \cite{wu2023personalized, wu2019nrms, wu2019npa}, whereas UEs create user-level representations by aggregating and contextualizing the embeddings of clicked news from the user's history \cite{okura2017embedding, an2019lstur, wu2022news}. The candidate's recommendation score is computed by comparing its embedding against the user embedding \cite{wang2018dkn, wu2019naml}. NNRs are primarily trained via point-wise classification objectives with negative sampling \cite{huang2013learning, wu2021empowering}. 
Exploiting users' past behavior as NNR supervision leads to recommendations that are content-wise closest to previously consumed news, in contrast to methods based on non-personalized criteria \cite{son2013location, chen2017location, ludmann2017recommending}.
More recent NNRs seek to augment content-based personalization by considering other aspects, such as categories, sentiment, emotions \cite{sertkan2022exploring}, entities \cite{iana2022survey}, outlets, or recency \cite{wu2023personalized}. These are incorporated in the NNR either as additional input to the NE \cite{wang2018dkn, gao2018fine, wu2019naml, liu2020kred, sheu2020context, lu2020beyond, qi2021personalized, xun2021we}, or in the form of an auxiliary training objective for the NE \cite{wu2019tanr, wu2020sentirec, qi2021pp}.

\vspace{1.4mm}
\noindent\textbf{Diversification.}
Personalized NNR reduces exposure to news dissimilar from those consumed in the past. Recommending ``more of the same'' constrains access to diverse viewpoints and information \cite{freedman1965selective,heitz2022benefits} and leads to homogeneous news diets and ``filter bubbles'' \cite{pariser2011filter}, in turn reinforcing users' initial stances \cite{li2019survey}.
Consequently, a significant body of work attempts to diversify recommendations, either by re-ranking them to increase some measure of diversity (e.g. intra-list distance \cite{zhang2008avoiding}) or by resorting to multi-task training \cite{gabriel2019contextual,wu2020sentirec,shi2022dcan,wu2022end,choi2022not,raza2023bias}, coupling the primary content-based personalization objective with auxiliary objectives that force aspect-based diversification.

\vspace{1.4mm}
\noindent\textbf{Current NNR Limitations.} 
Critically, existing approaches, by ``hardcoding'' aspectual requirements (i.e., personalization or diversification for an aspect) into the NNR's architecture and/or training objectives, cannot be easily adjusted for varying recommendation goals. Since even minor changes in the recommendation objective require retraining the NNR,  current models are generally limited to fixed single-aspect recommendation scenarios (e.g., content-based personalization with topical diversification), and ill-equipped for multi-aspect customization.  
In this work, we rethink personalized news recommendation and propose a novel, modular multi-aspect recommendation framework that allows for ad-hoc creation of recommendation functions over aspects at inference time. 
This enables fundamentally different recommendation: one that lets each user define their own custom recommendation function, choosing the amount of personalization or diversification for each aspect.

\section{Methodology}
\label{sec:methodology}

\begin{figure*}[t]
    \centering
    \includegraphics[width=\textwidth]{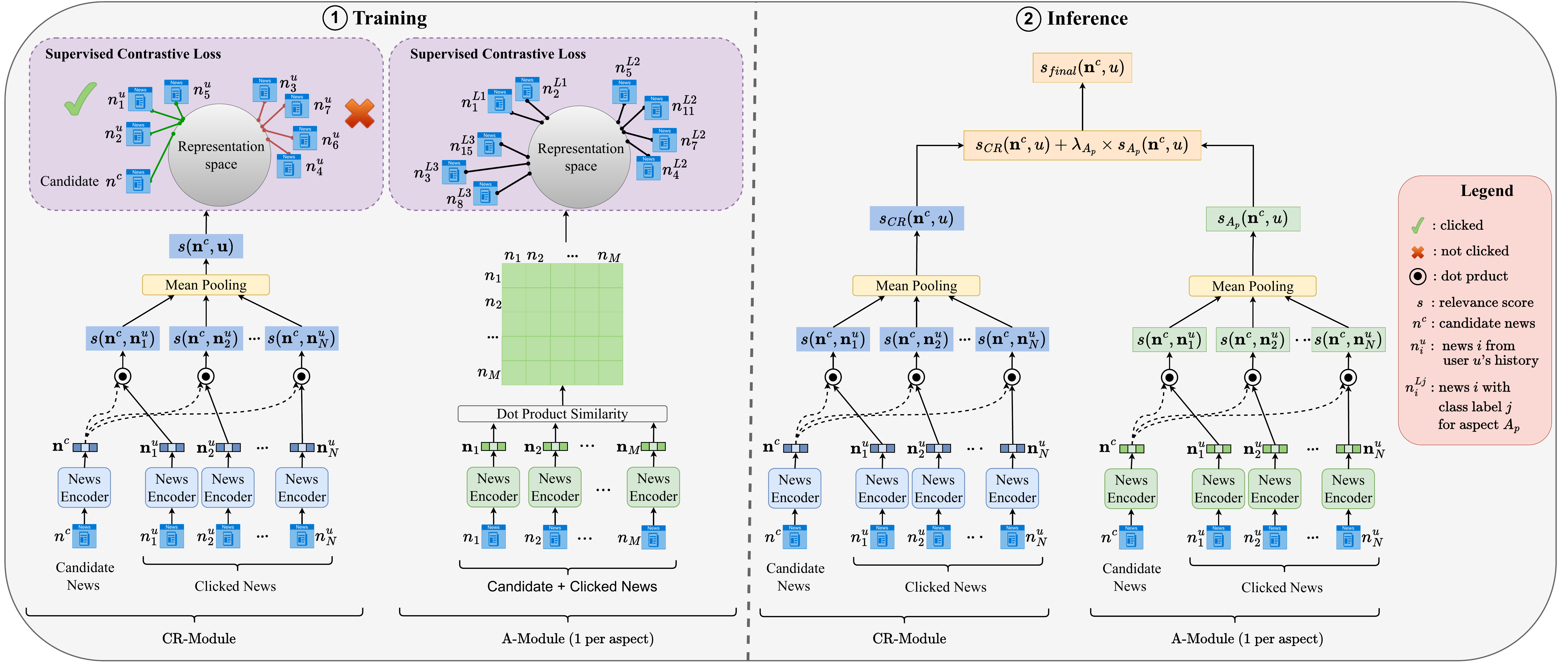}
    \caption{Illustration of the \manner{} framework. \textcircled{\raisebox{-.9pt} {1}} We train aspect-specialized NEs (i.e. \texttt{CR-Module} for content-based personalization, \texttt{A-Module} for aspect-based similarity) with metric-based contrastive learning. \textcircled{\raisebox{-.9pt} {2}} Inference: we linearly aggregate aspect-specific similarity scores between candidate and clicked news for final ranking.}
    \label{fig:framework}
    \vspace{-0.5em}
\end{figure*}

Personalized news recommendation produces for each candidate news $n^c$ and user $u$ with corresponding click history $H \hspace{-0.2em} = \hspace{-0.2em} \{ n^u_1, n^u_2, ..., n^u_N \} $, a relevance score $s(n^c, u)$ that quantifies the candidate's relevance for the user. 
We define an \textit{aspect} $A_p$ as a categorical variable that encodes a news attribute (e.g. its category, stance, sentiment, provider), where each news $n_i$ can belong only to one value of $A_p$ (e.g. if $A_p$ is the topic, then $n_i$ may take exactly one value from \{\textit{politics}, \textit{sports}, ...\}).
As discussed in \S\ref{sec:related_work}, aspects are additional dimensions next to content over which to tailor recommendations, whether by (i) personalizing or (ii) diversifying over them. In line with earlier work, we define \textit{aspect-based personalization} as the level of homogeneity between a user's recommendations and clicked news w.r.t. the distribution of aspect $A_p$. In contrast, we define \textit{aspect-based diversity} as the level of uniformity of aspect $A_p$'s 
distribution among the news in the recommendation list. 

We next introduce our proposed \textit{modular framework} \manner{}, illustrated in Fig. \ref{fig:framework}. Starting from a PLM, during (1) training, we reshape the PLM's representation space via contrastive learning, independently for each aspect; this results in one specialized NE for each aspect; at (2) inference, we can, depending on the recommendation task, aggregate the resulting aspect-specific similarity scores to produce a final ranking function.

\subsection{News Encoder}
\label{subsec:news_encoder}

We adopt a dual-component architecture for the NE coupling (i) a text and (ii) an entity encoder \cite{qi2021pp,qi2021hierec}. 
The former, a PLM, transforms the text input (i.e., concatenation of news title and abstract) into a text-based news embedding $\mathbf{n}_{t}$, given by the PLM's output \texttt{[CLS]} token representation. 
The latter learns an entity-level news embedding $\mathbf{n}_e$ by contextualizing pretrained embeddings of named entities (i.e., extracted from title and abstract) in a layer that combines multi-head self-attention \cite{vaswani2017attention} and additive attention \cite{bahdanau2014neural}. 
The final news embedding $\mathbf{n}$ is the concatenation of $\mathbf{n}_t$ and $\mathbf{n}_e$.

\subsection{Modular Training}
\label{subsec:training}

\manner{} comprises two module types, each with a dedicated NE, responsible for content-based (\texttt{CR-Module)} and aspect-based (\texttt{A-Module}) recommendation relevance, respectively. 
We train both by minimizing the supervised contrastive loss (SCL, Eq.~\ref{eq:scl}) which aims to reshape the NE's representation space so that embeddings of same-class instances become mutually closer (cf. a distance/similarity metric) than instances of different classes \cite{khosla2020supervised,gunelsupervised}. To this end, we contrast the similarity score of a positive example (pair of same-class instances) against scores of corresponding negative examples (paired instances from different classes):
\begin{equation}
\small
  \mathcal{L} \hspace{-0.2em} = 
  \hspace{-0.2em}-\hspace{-0.4em}\sum_{i=1}^N\frac{1}{N_{y_i} -1} 
  \hspace{-0.8em}
  \sum_{\substack{
      j \in [1, N] \\
      i \neq j, y_i=y_j
      }}
  \hspace{-1em}
  \log 
  \frac{
    e^{(\mathbf{n}_i \cdot \mathbf{n}_j / \tau)}
    }
    {
    \sum_{\substack{
            k \in [1, N] \\ 
            i \neq k 
            }}
    e^{(\mathbf{n}_i \cdot \mathbf{n}_k / \tau)}
    }
  \label{eq:scl}
\end{equation}
with $y_i$ as news $n_i$'s label, $N$ the batch size, $N_{y_i}$ the number of batch instances with label $y_i$, and $\tau \hspace{-0.2em} > \hspace{-0.2em} 0$ the temperature hyperparameter controlling the extent of class separation. 
We use the dot product as the similarity metric for both module types. 

\vspace{1.4mm}
\noindent\textbf{CR-Module.}
Our \texttt{CR-Module} is a modification of the common content-based NNR architecture \cite{wu2023personalized}. Concretely, we encode both candidate and clicked news with a dedicated NE. However, following \citet{iana2023simplifying}, we replace the widely used UEs (i.e., early fusion of clicked news representations) with the simpler (and non-parameterized) mean-pooling of dot-product scores between the candidate embedding $\mathbf{n}^c$ and clicked news embeddings $\mathbf{n}_i^u$: $s(\mathbf{n}^c, u) \hspace{-0.2em} = \hspace{-0.2em} \frac{1}{N} \sum_{i=1}^N \mathbf{n}^c \hspace{-0.2em} \cdot \hspace{-0.2em} \mathbf{n}_i^u$ (i.e., late-fusion).
We thus reduce the computational complexity of the standard approaches with elaborate parameterized UEs. We then update the \texttt{CR-Module}'s encoder (i.e., fine-tune the PLM) by minimizing SCL, with clicked candidates as positive and non-clicked news as negative examples for the user. As there are many more non-clicked news, we resort to negative sampling \cite{ijcai2022infonce}.

\vspace{1.4mm}
\noindent\textbf{A-Module.}
Each \texttt{A-Module} trains a specialized NE for one aspect other than content. Via the metric-based objective, we reshape the PLM's representation space to group news according to aspect classes. Given a multi-class aspect, we first construct the training set from the union of all news in the dataset. Sets of news with the same aspect label form the positive samples for SCL; we obtain the corresponding negatives by pairing the same news from positive pairs with news from other aspect classes (e.g., for topical category as $A_p$, a news from \textit{sports} is paired with the news from \textit{politics} and/or \textit{weather}). 
For each aspect, we independently fine-tune a separate copy of the same initial PLM. 
Note that the resulting aspect-specific NE encodes no information on user preferences: it only encodes the news similarity w.r.t. the aspect in question. Importantly, this implies that extending \manner{} to support a new aspect amounts to merely training an additional \texttt{A-Module} for that aspect.

\subsection{Inference: Custom Ranking Functions}
\label{subsec:inference}

At inference time, the NEs of the \texttt{CR-Module} and of each of the \texttt{A-Modules} are leveraged identically: we encode the candidate news as well as the user's clicked news with the respective NE, obtaining their module-specific embeddings $\mathbf{n}^c$ and $\mathbf{n}^{u}_i$ -- their dot product $s \hspace{-0.2em} = \hspace{-0.2em} \mathbf{n}^c \hspace{-0.2em} \cdot \hspace{-0.2em} \mathbf{n}^{u}_i$ quantifies their similarity according to the module's aspect (or content for \texttt{CR-Module}'s NE).  
As different NEs produce similarity scores of different  magnitudes, we z-score normalize each module's scores per user. The final ranking score constitutes a \textit{linear} aggregation of the content $s_{CR}$ and aspect $s_{A_p}$ similarity scores: 
\begin{equation}
\small
s_{final}(\mathbf{n}^c, \mathbf{u}) \hspace{-0.2em} = \hspace{-0.2em} s_{CR} + \hspace{-0.4em} \sum_{A_p \in A} \lambda_{A_p} s_{A_p}
\end{equation}
\noindent where $\lambda_{A_p}$ is the scaling weight for the aspect score, and $A$ the set of all aspects of interest. 
This linear composability of aspect-specific similarity scores allows not only generalization to multi-aspect recommendation objectives, but also different ad-hoc realizations of the ranking function that match custom recommendation goals: (i) with $\lambda_{A_p} \hspace{-0.2em} = \hspace{-0.2em} 0$, \manner{} performs standard content-based personalization, (ii) for $\lambda_{A_p} \hspace{-0.2em} > \hspace{-0.2em} 0$ it recommends based on both content- and aspect personalization, whereas (iii) for $\lambda_{A_p} \hspace{-0.2em} < \hspace{-0.2em} 0$ it simultaneously personalizes by content but diversifies for the aspect(s). 

\section{Experimental Setup}
\label{sec:experimental_setup}

We compare \manner{} against state-of-the-art NNRs on a range of single- and multi-aspect recommendation tasks. We experiment with two aspects: \textit{topical categories} (\textit{ctg}) and news \textit{sentiment} (\textit{snt}).

\vspace{1.4mm}
\noindent\textbf{Baselines.} 
We evaluate several NNRs trained on classification objectives. We follow \citet{wu2021empowering} and replace the original NEs of all baselines that do not use PLMs (instead, contextualizing word embeddings with convolution or self-attention layers) with the same PLM used in \manner{}.\footnote{The only exception is the final text embedding, where \citet{wu2021empowering} pool tokens with an attention network.} 
We include two models optimized purely for content personalization: (1) NRMS \cite{wu2019nrms}, and (2) MINER \cite{li2022miner}. We further evaluate seven NNRs that inject aspect information. Thereof, five incorporate \textit{topical categories}, i.e., (3) NAML \cite{wu2019naml}, (4) LSTUR \cite{an2019lstur}, (5) MINS \cite{wang2022news}, (6) CAUM \cite{qi2022news}, (7) TANR \cite{wu2019tanr}, and two the news \textit{sentiment}: (8) SentiRec \cite{wu2020sentirec}, and (9) SentiDebias \cite{wu2022removing}. 

\vspace{1.4mm}
\noindent\textbf{Data.}
We carry out the evaluation on two prominent monolingual news recommendation benchmarks: MINDlarge (denoted MIND) \cite{wu2020mind} with news in English and Adressa-1 week \cite{gulla2017adressa} (denoted Adressa) with Norwegian news. 
Since \citet{wu2020mind} do not release test labels for MIND, we use the provided validation portion for testing, and split the respective training set into temporally disjoint training (first four days of data) and validation portions (the last day). Following established practices on splitting the Adressa dataset \cite{hu2020graph,xu2023group}, we use the data of the first five days to construct user histories and the clicks of the sixth day to build the training dataset. We randomly sample 20\% of the last day's clicks to create the validation set, and treat the remaining samples of the last day as the test set.\footnote{Note that during validation and testing, we reconstruct user histories with all the samples of the first six days of data.} Since Adressa contains only positive samples (i.e., no data about users' seen but not clicked news), we randomly sample 20 news as negatives for each clicked article to build impressions following \citet{yi2021efficient}.\footnote{Table \ref{tab:datasets} summarizes the datasets' statistics.}
As Adressa contains no disambiguated named entities, we use only the news title as input to \manner{}' NE, while on MIND we use all news features as NE input.

Regarding aspects, the topical category annotations are provided in both datasets. As no sentiment labels exist in neither MIND nor Adressa, we use a multilingual XLM-RoBERTa Base model \cite{conneau2020unsupervised} trained on tweets and fine-tuned for sentiment analysis \cite{barbieri2022xlm} to classify news into three classes: positive (pos), neutral, and negative (neg). We compute real-valued scores using the model's confidence scores $s_i$ for class $i$, and the predicted sentiment class label $\hat{l}$ as follows:
\begin{equation}
\small
    s_{sent}  \hspace{-0.2em} =  \hspace{-0.2em}
        \begin{cases}
            (+1) \times s_{pos} \text{, if } \hat{l} \hspace{-0.2em} =  \hspace{-0.2em} pos \\  
            (-1) \times s_{neg} \text{, if } \hat{l} \hspace{-0.2em} =  \hspace{-0.2em} neg \\  
            (1 - s_{neutral})  \hspace{-0.2em} \times \hspace{-0.2em} (s_{pos} - s_{neg}) \text{, otherwise} \\  
        \end{cases}
\end{equation}

\vspace{1.4mm}
\noindent\textbf{Evaluation Metrics.}
We report performance with AUC, MRR, nDCG@k ($k \hspace{-0.2em} = \hspace{-0.2em} \{5, 10\}$). We measure aspect-based diversity of recommendations at position $k$ as the normalized entropy of the distribution of aspect $A_p$'s values in the recommendation list: 
\begin{equation}
\small
    D_{A_p}@k \hspace{-0.2em} = \hspace{-0.2em} - \sum_{j \in A_p} \frac{p(j) \log p(j)}{\log(|A_p|)}
\end{equation}
where $A_p \hspace{-0.2em} \in \hspace{-0.2em} \{ctg, snt\}$, and $|A_p|$ is the number of $A_p$ classes.
If aspect-based personalization is successful, aspect $A_p$'s distribution in the recommendations should be similar to its distribution in the user history. We evaluate personalization with the generalized Jaccard similarity \cite{bonnici2020kullback}:
\begin{equation}
\small
    \mathit{PS}_{A_p}@k \hspace{-0.2em} = \hspace{-0.2em} \frac{\sum_{j=1}^{|A_p|} \min(\mathcal{R}_j, \mathcal{H}_j)}{\sum_{j=1}^{|A_p|} \max(\mathcal{R}_j, \mathcal{H}_j)},
\end{equation}
where $R_j$ and $H_j$ represent the probability of a news with class $j$ of $A_p$ to be contained in the recommendations list $R$, and, respectively, in the user history $H$.
As all metrics are bounded to $[0, 1]$, we measure the trade-off between content-based personalization (nDCG@$k$) and either aspect-based diversity $D_{A_p}@k$ or aspect-based personalization $\mathit{PS}_{A_p}@k$ with the harmonic mean. We denote this T\textsubscript{A\textsubscript{p}}@$k$ for single-aspect. For multi-aspect evaluation, i.e., when ranking for content-personalization by diversifying simultaneously over topics and sentiment, we adopt as evaluation metric the harmonic mean between nDCG@$k$, D\textsubscript{ctg}@$k$ (topical category), and D\textsubscript{snt}@$k$ (sentiment), denoted 
T\textsubscript{all}@$k$. 

\vspace{1.4mm}
\noindent\textbf{Training Details.}
We use RoBERTa Base \cite{liu2019roberta} and NB-BERT Base \cite{kummervold2021operationalizing,nielsen2023scandeval} in experiments on MIND and Adressa, respectively. 
We set the maximum history length to 50. We tune the main hyperparameters of all NNRs. We train all models with mixed precision, the Adam optimizer \cite{kingma2014adam}, the learning rate of 1e-5 on MIND, 1e-6 on Adressa, and 1e-6 for the sentiment\,\texttt{A-Module} on both datasets.
In \texttt{A-Module} training, we sample 20 instances per class,\footnote{For $M$ class instances, we obtain $\frac{M^2-M}{2}$ positive pairs for that class for SCL. 
} while in \texttt{CR-Module} training we sample four negatives per positive example. We find the optimal temperature of 0.36 on MIND, and 0.14 on Adressa, for the \texttt{CR-Module}, and of 0.9 for all \texttt{A-Modules} on both datasets.
We train all baselines and the \texttt{CR-Module} for 5 epochs on MIND and 20 epochs on Adressa, with a batch size of 8. We train each \texttt{A-Module} for 100 epochs, with the batch size of 60 for sentiment and 360 for topics. We repeat runs five times with different seeds and report averages and standard deviations for all metrics. We refer to Appendices \ref{sec:appendix_model_parameters} - \ref{sec:appendix_hyperparameters} for further details about model sizes and hyperparameters.

\section{Results and Discussion}
\label{sec:results_discussion}
\begin{table*}[ht]
\centering
\scriptsize
\resizebox{\textwidth}{!}{%
    \begin{tabular}{l|cccc|cccc}
        \toprule
        \multicolumn{1}{c}{} & \multicolumn{4}{c}{\textbf{MIND}} & \multicolumn{4}{c}{\textbf{Adressa}} \\  
        \cmidrule(lr){2-5} \cmidrule(lr){6-9}
        
        \multicolumn{1}{l|}{Model} 
        & AUC & MRR & nDCG@5 & nDCG@10
        & AUC & MRR & nDCG@5 & nDCG@10
        \\
        \hline

       NRMS-PLM 
        & 63.0\textsubscript{$\pm$1.5} 
        & 35.5\textsubscript{$\pm$0.6}  
        & 33.4\textsubscript{$\pm$0.7} 
        & 39.9\textsubscript{$\pm$0.6}  

        & 72.3\textsubscript{$\pm$3.3} 
        & 43.0\textsubscript{$\pm$2.7} 
        & 44.3\textsubscript{$\pm$2.8} 
        & 51.3\textsubscript{$\pm$2.3} 
        \\

        MINER
        & 63.1\textsubscript{$\pm$1.2} 
        & 35.5\textsubscript{$\pm$1.1}  
        & 33.7\textsubscript{$\pm$1.1} 
        & 40.0\textsubscript{$\pm$1.0}  

        & 70.1\textsubscript{$\pm$4.9} 
        & 37.3\textsubscript{$\pm$4.1} 
        & 38.5\textsubscript{$\pm$5.1} 
        & 46.3\textsubscript{$\pm$4.1} 
        \\

        \hdashline

        NAML-PLM
        & 60.6\textsubscript{$\pm$3.4} 
        &  \underline{37.6\textsubscript{$\pm$0.4}}  
        &  \underline{35.9\textsubscript{$\pm$0.4}} 
        &  \underline{42.2\textsubscript{$\pm$0.4}}  

        & 50.0\textsubscript{$\pm$0.0} 
        & \underline{45.0\textsubscript{$\pm$5.0}} 
        & 47.2\textsubscript{$\pm$5.5} 
        & 52.5\textsubscript{$\pm$4.1} 
        \\

        LSTUR-PLM
        & 54.6\textsubscript{$\pm$3.0} 
        & 33.3\textsubscript{$\pm$1.5}  
        & 31.7\textsubscript{$\pm$1.8} 
        & 38.3\textsubscript{$\pm$1.7}  

        & 65.0\textsubscript{$\pm$7.2} 
        & 43.1\textsubscript{$\pm$1.7} 
        & 44.8\textsubscript{$\pm$2.6} 
        & 51.2\textsubscript{$\pm$2.0} 
        \\

        MINS-PLM 
        & 61.3\textsubscript{$\pm$2.7} 
        & 36.2\textsubscript{$\pm$0.3}  
        & 34.5\textsubscript{$\pm$0.4} 
        & 40.8\textsubscript{$\pm$0.3}  

        & 74.3\textsubscript{$\pm$3.2} 
        & 44.2\textsubscript{$\pm$2.9} 
        & \underline{47.3\textsubscript{$\pm$3.3}} 
        & \underline{53.0\textsubscript{$\pm$3.4}} 
        \\ 

        CAUM\textsubscript{no entities}-PLM
        &  \underline{66.2\textsubscript{$\pm$3.0}} 
        & 36.6\textsubscript{$\pm$2.0}  
        & 34.6\textsubscript{$\pm$2.0} 
        & 41.0\textsubscript{$\pm$1.9}  

        & \underline{76.5\textsubscript{$\pm$1.2}} 
        & 43.6\textsubscript{$\pm$1.3} 
        & 46.9\textsubscript{$\pm$1.3} 
        & 52.0\textsubscript{$\pm$1.2} 
        \\

        CAUM-PLM 
        & 66.4\textsubscript{$\pm$3.1} 
        & 36.2\textsubscript{$\pm$1.2}  
        & 34.3\textsubscript{$\pm$1.3} 
        & 40.8\textsubscript{$\pm$1.3}  

        & -- 
        & -- 
        & -- 
        & -- 
        \\

        TANR-PLM
        & 63.3\textsubscript{$\pm$1.1} 
        & 37.0\textsubscript{$\pm$1.0}  
        & 35.2\textsubscript{$\pm$1.0} 
        & 41.6\textsubscript{$\pm$0.9}  

        & 50.0\textsubscript{$\pm$0.0} 
        & 43.8\textsubscript{$\pm$1.0} 
        & 45.6\textsubscript{$\pm$1.3} 
        & 51.4\textsubscript{$\pm$0.6} 
        \\
        
        \hdashline

        SentiRec-PLM
        & 62.2\textsubscript{$\pm$0.7} 
        & 35.7\textsubscript{$\pm$0.4}  
        & 33.9\textsubscript{$\pm$0.4} 
        & 40.5\textsubscript{$\pm$0.4}  

        & 67.6\textsubscript{$\pm$2.7} 
        & 33.1\textsubscript{$\pm$2.4} 
        & 32.9\textsubscript{$\pm$3.8} 
        & 40.8\textsubscript{$\pm$2.4} 
        \\

        SentiDebias-PLM
        & 55.0\textsubscript{$\pm$2.5} 
        & 27.8\textsubscript{$\pm$1.9}  
        & 25.5\textsubscript{$\pm$1.9} 
        & 32.2\textsubscript{$\pm$2.0}  

        & 67.4\textsubscript{$\pm$2.4} 
        & 35.7\textsubscript{$\pm$3.4}  
        & 36.4\textsubscript{$\pm$4.2} 
        & 44.2\textsubscript{$\pm$2.9}  
        \\ 
        \hline
        
        \manner{} (\texttt{CR-Module})
        & \textbf{69.7\textsubscript{$\pm$0.9 }} 
        & \textbf{38.6\textsubscript{$\pm$0.6}}  
        & \textbf{37.0\textsubscript{$\pm$0.6}} 
        & \textbf{43.2\textsubscript{$\pm$0.6}}  

        & \textbf{79.4\textsubscript{$\pm$1.7}} 
        & \textbf{47.0\textsubscript{$\pm$2.4}} 
        & \textbf{48.9\textsubscript{$\pm$2.8}} 
        & \textbf{54.3\textsubscript{$\pm$2.5}} 
        \\
        \hline

        \rowcolor{Gray}
        Improvement (\%)
        & + 5.4
        & + 2.8
        & + 3.1
        & + 2.3

        & + 3.7
        & + 4.6
        & + 3.3
        & + 2.5
        \\ 
        
        \bottomrule
        
    \end{tabular}%
    }
\caption{Content-based recommendation performance. We average results across five runs, and report the relative improvement over the best baseline. The best results per column are highlighted in bold, the second best underlined.}
\label{tab:results_recommendation}

\vspace{-0.5em}
\end{table*}

\newcolumntype{g}{>{\columncolor{Gray}}c}

\begin{table*}[ht]
\centering

\resizebox{\textwidth}{!}{%
    \begin{tabular}{l|cgcgcg|cgcgcg}
        \toprule
        \multicolumn{1}{c}{} & \multicolumn{6}{c}{\textbf{MIND}} & \multicolumn{6}{c}{\textbf{Adressa}} \\  
        \cmidrule(lr){2-7} \cmidrule(lr){8-13}
        
        \multicolumn{1}{l|}{Model} 
        & nDCG@10 
        & D\textsubscript{ctg}@10 & T\textsubscript{ctg}@10 
        & D\textsubscript{snt}@10 & T\textsubscript{snt}@10 
        & T\textsubscript{all}@10
        
        & nDCG@10 
        & D\textsubscript{ctg}@10 & T\textsubscript{ctg}@10 
        & D\textsubscript{snt}@10 & T\textsubscript{snt}@10 
        & T\textsubscript{all}@10
        \\
        \hline

       NRMS-PLM 
        & 39.9\textsubscript{$\pm$0.6} 
        & 50.0\textsubscript{$\pm$1.1}  
        & 44.3\textsubscript{$\pm$0.4}  
        & 66.4\textsubscript{$\pm$0.5}  
        & 49.8\textsubscript{$\pm$0.5}  
        & 49.9\textsubscript{$\pm$0.3}  

        & 51.3\textsubscript{$\pm$2.3}  
        & 31.8\textsubscript{$\pm$1.0}  
        & 39.2\textsubscript{$\pm$0.5}  
        & 61.5\textsubscript{$\pm$0.5}  
        & 55.9\textsubscript{$\pm$1.2}  
        & 44.6\textsubscript{$\pm$0.5}  
        \\

        MINER
        & 40.0\textsubscript{$\pm$1.0} 
        & 49.4\textsubscript{$\pm$1.2} 
        & 44.2\textsubscript{$\pm$0.4} 
        & 65.7\textsubscript{$\pm$0.9} 
        & 49.7\textsubscript{$\pm$1.0} 
        & 49.6\textsubscript{$\pm$0.5} 

        & 46.3\textsubscript{$\pm$4.1}  
        & 31.1\textsubscript{$\pm$0.6}  
        & 37.1\textsubscript{$\pm$1.6}  
        & 60.9\textsubscript{$\pm$0.5}  
        & 52.5\textsubscript{$\pm$2.8}  
        & 42.7\textsubscript{$\pm$1.5}  
        \\
        \hdashline

         NAML-PLM
        & 42.2\textsubscript{$\pm$0.4} 
        & 47.3\textsubscript{$\pm$0.3}  
        & 44.6\textsubscript{$\pm$0.3}  
        & 65.1\textsubscript{$\pm$0.4}  
        & 51.2\textsubscript{$\pm$0.3}  
        & 49.9\textsubscript{$\pm$0.3}  

        & 52.5\textsubscript{$\pm$4.1}  
        & 30.6\textsubscript{$\pm$2.4}  
        & 38.6\textsubscript{$\pm$2.1}  
        & 61.6\textsubscript{$\pm$0.6}  
        & 56.7\textsubscript{$\pm$2.6}  
        & 44.0\textsubscript{$\pm$1.9}  
        \\

        LSTUR-PLM
        & 38.3\textsubscript{$\pm$1.7} 
        & 50.0\textsubscript{$\pm$1.2}  
        & 43.4\textsubscript{$\pm$0.7}  
        & 65.6\textsubscript{$\pm$0.3}  
        & 48.4\textsubscript{$\pm$1.3}  
        & 48.9\textsubscript{$\pm$0.5}  

        & 51.2\textsubscript{$\pm$2.0}  
        & 29.9\textsubscript{$\pm$4.6}  
        & 37.7\textsubscript{$\pm$5.2}  
        & 61.4\textsubscript{$\pm$0.5}  
        & 55.8\textsubscript{$\pm$1.2}  
        & 43.2\textsubscript{$\pm$3.8}  
        \\

        MINS-PLM
        & 40.8\textsubscript{$\pm$0.3} 
        & 49.1\textsubscript{$\pm$1.0}  
        & 44.6\textsubscript{$\pm$0.3}  
        & 66.3\textsubscript{$\pm$0.9}  
        & 50.5\textsubscript{$\pm$0.1}  
        & 50.0\textsubscript{$\pm$0.4}  

        & 53.0\textsubscript{$\pm$3.4}  
        & 33.6\textsubscript{$\pm$1.7}  
        & \underline{41.0\textsubscript{$\pm$1.0}}  
        & 61.8\textsubscript{$\pm$0.6}  
        & 57.0\textsubscript{$\pm$1.8}  
        & \underline{46.2\textsubscript{$\pm$0.9}}  
        \\

        CAUM\textsubscript{no entities}-PLM
        & 41.0\textsubscript{$\pm$1.9} 
        & 47.4\textsubscript{$\pm$1.0}  
        & 43.9\textsubscript{$\pm$0.9}  
        & 65.8\textsubscript{$\pm$1.2}  
        & 50.5\textsubscript{$\pm$1.3}  
        & 49.4\textsubscript{$\pm$0.6}  

        & 52.0\textsubscript{$\pm$1.2}  
        & \underline{34.4\textsubscript{$\pm$0.3}}  
        & \textbf{41.4\textsubscript{$\pm$0.4}}  
        & 62.1\textsubscript{$\pm$0.5}  
        & 56.6\textsubscript{$\pm$0.7}  
        & \textbf{46.6\textsubscript{$\pm$0.3}}  
        \\

        CAUM-PLM
        & 40.8\textsubscript{$\pm$1.3} 
        & 47.8\textsubscript{$\pm$0.9}  
        & 44.0\textsubscript{$\pm$1.0}  
        & 66.1\textsubscript{$\pm$0.5}  
        & 50.6\textsubscript{$\pm$1.0}  
        & 49.6\textsubscript{$\pm$0.9}  

        & --  
        & --  
        & --  
        & --  
        & --  
        & --  
        \\

        TANR-PLM 
        & 41.6\textsubscript{$\pm$0.9} 
        & 48.9\textsubscript{$\pm$0.9}  
        & 45.0\textsubscript{$\pm$0.3}  
        & 66.1\textsubscript{$\pm$0.8}  
        & 51.1\textsubscript{$\pm$0.7}  
        & 50.3\textsubscript{$\pm$0.3}  

        & 51.4\textsubscript{$\pm$0.6}  
        & 32.9\textsubscript{$\pm$1.7}  
        & 40.1\textsubscript{$\pm$1.1}  
        & 61.8\textsubscript{$\pm$0.7}  
        & 56.1\textsubscript{$\pm$0.2}  
        & 45.4\textsubscript{$\pm$1.0}  
        \\
        \hdashline

        SentiRec-PLM
        & 40.5\textsubscript{$\pm$0.4} 
        & 49.4\textsubscript{$\pm$0.4}  
        & 44.5\textsubscript{$\pm$0.1}  
        & 67.0\textsubscript{$\pm$0.6}  
        & 50.4\textsubscript{$\pm$0.4}  
        & 50.1\textsubscript{$\pm$0.2}  

        & 40.8\textsubscript{$\pm$2.4}  
        & \textbf{35.6\textsubscript{$\pm$0.6}}  
        & 38.0\textsubscript{$\pm$1.1}  
        & \textbf{68.5\textsubscript{$\pm$0.2}}  
        & 51.1\textsubscript{$\pm$1.9}  
        & 44.6\textsubscript{$\pm$1.0}  
        \\ 

        SentiDebias-PLM
        & 32.2\textsubscript{$\pm$2.0} 
        & \textbf{52.0\textsubscript{$\pm$2.2}}  
        & 39.7\textsubscript{$\pm$1.1}  
        & \underline{68.6\textsubscript{$\pm$1.2}}  
        & 43.8\textsubscript{$\pm$1.8}  
        & 46.2\textsubscript{$\pm$1.0}  

        & 44.2\textsubscript{$\pm$2.9} 
        & 32.3\textsubscript{$\pm$1.0}  
        & 37.3\textsubscript{$\pm$1.2}  
        & 61.2\textsubscript{$\pm$0.2}  
        & 51.3\textsubscript{$\pm$2.0}  
        & 42.9\textsubscript{$\pm$1.1}  
        \\
        \hline

        \manner{} (\texttt{CR-Module})
        & \textbf{43.2\textsubscript{$\pm$0.6}} 
        & 49.3\textsubscript{$\pm$0.3}  
        & \underline{46.0\textsubscript{$\pm$0.3}}  
        & 65.4\textsubscript{$\pm$0.6}  
        & \underline{52.0\textsubscript{$\pm$0.4}}  
        & 51.1\textsubscript{$\pm$0.2}  

        & \textbf{54.3\textsubscript{$\pm$2.5}}  
        & 31.7\textsubscript{$\pm$0.2}  
        & 40.0\textsubscript{$\pm$0.7}  
        & 61.4\textsubscript{$\pm$0.3}  
        & \underline{57.6\textsubscript{$\pm$1.5}}  
        & 45.3\textsubscript{$\pm$0.6}  
        \\ 

        \manner{} ($\lambda_{ctg}=-0.2 / -0.3$, $\lambda_{snt}=0$)
        & 42.0\textsubscript{$\pm$0.6} 
        & \underline{51.5\textsubscript{$\pm$0.3}}  
        & \textbf{46.2\textsubscript{$\pm$0.3}}  
        & 65.6\textsubscript{$\pm$0.6}  
        & 51.2\textsubscript{$\pm$0.4}  
        & \underline{51.3\textsubscript{$\pm$0.3}}  

        & 50.9\textsubscript{$\pm$2.5}  
        & 34.1\textsubscript{$\pm$0.3}  
        & 40.8\textsubscript{$\pm$0.8}  
        & 61.9\textsubscript{$\pm$0.3}  
        & 55.8\textsubscript{$\pm$1.6}  
        & 46.0\textsubscript{$\pm$0.7}  
        \\ 

        \manner{} ($\lambda_{ctg}=0$, $\lambda_{snt}=-0.3 / -0.2$)
        & \underline{42.8\textsubscript{$\pm$0.7}} 
        & 49.8\textsubscript{$\pm$0.2}  
        & 46.0\textsubscript{$\pm$0.4}  
        & \textbf{68.7\textsubscript{$\pm$0.3}}  
        & \textbf{52.7\textsubscript{$\pm$0.4}}  
        & \textbf{51.7\textsubscript{$\pm$0.3}}  

        & \underline{53.8\textsubscript{$\pm$2.5}}  
        & 32.4\textsubscript{$\pm$0.2}  
        & 40.4\textsubscript{$\pm$0.7}  
        & \underline{63.0\textsubscript{$\pm$0.3}}  
        & \textbf{58.0\textsubscript{$\pm$1.5}}  
        & 45.9\textsubscript{$\pm$0.6}  
        \\ 
        
        \bottomrule
        
    \end{tabular}%
    }
\caption{Single-aspect diversification. For \manner{}, we list the best results (cf. T\textsubscript{A\textsubscript{p}}) of single-aspect diversification as  $\lambda_{A_p}$ (MIND/Adressa). The best results per column are highlighted in bold, the second best underlined.}
\label{tab:results_diversification}

\vspace{-0.5em}
\end{table*}

We first discuss \manner{}'s content personalization performance. We then analyze its capability for single- and multi-aspect (i) diversification and (ii) personalization. In the aspect customization setups, we treat \manner{}'s \texttt{CR-Module} as a baseline. Lastly, we evaluate its ability to re-use pretrained aspect-specific modules in cross-lingual transfer. 

\subsection{Content Personalization}
\label{sec:content_personalization}

Table \ref{tab:results_recommendation} summarizes the results on content personalization. Since the task does not require any aspect-based customization, we evaluate the \manner{} variant that uses only its CR-Module at inference time (i.e., $\lambda \hspace{-0.2em} = \hspace{-0.2em} 0$).
\manner{} consistently outperforms all state-of-the-art NNRs in terms of both classification and ranking metrics on both datasets. Given that \manner{}'s \texttt{CR-Module} derives the user embedding by merely averaging clicked news embeddings, these results question the need for complex parameterized UEs, present in all the baselines, in line with the findings of \citet{iana2023simplifying}. 

We ablate the \texttt{CR-Module}'s content personalization performance for (i) different inputs to the NE and (ii) alternative architecture designs and training objectives.
We find that all groups of features (e.g., abstract, named entities) contribute to the overall performance (cf. Fig. \ref{fig:ablation_features}). 
Moreover, we confirm the findings of \citet{iana2023simplifying} that (i) late fusion outperforms a parameterized UE (i.e., early fusion), and that (ii) SCL better separates classes than cross-entropy loss, in line with other similarity-oriented NLP tasks  \cite{reimers2019sentence}.

\subsection{Single-Aspect Customization}

\begin{figure*}[h]
     \centering
     \begin{subfigure}[b]{0.48\textwidth}
         \centering
         \includegraphics[width=\textwidth]{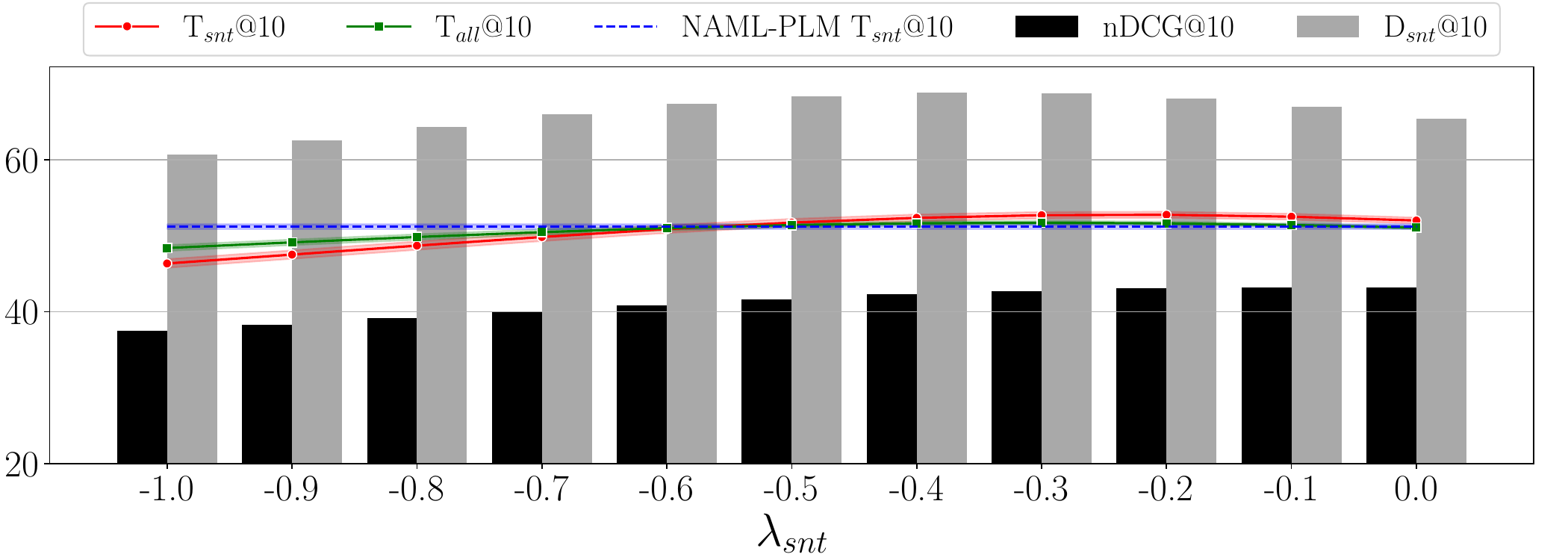}
         \caption{Single-aspect sentiment diversification.}
         \label{fig:single_aspect_div_sent_mind}
     \end{subfigure}
     \hfill
     \begin{subfigure}[b]{0.48\textwidth}
         \centering
         \includegraphics[width=\textwidth]{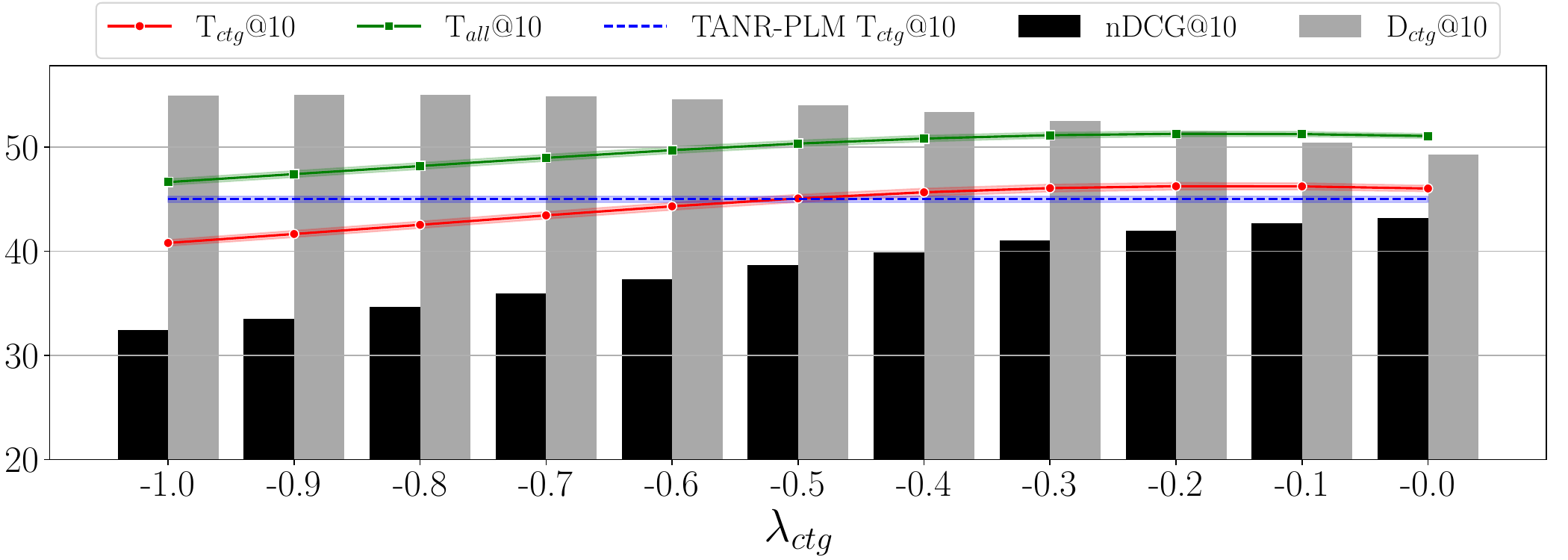}
         \caption{Single-aspect category diversification.}
         \label{fig:single_aspect_div_categ_mind}
     \end{subfigure}
     \hfill
     \begin{subfigure}[b]{0.48\textwidth}
         \centering
         \includegraphics[width=\textwidth]{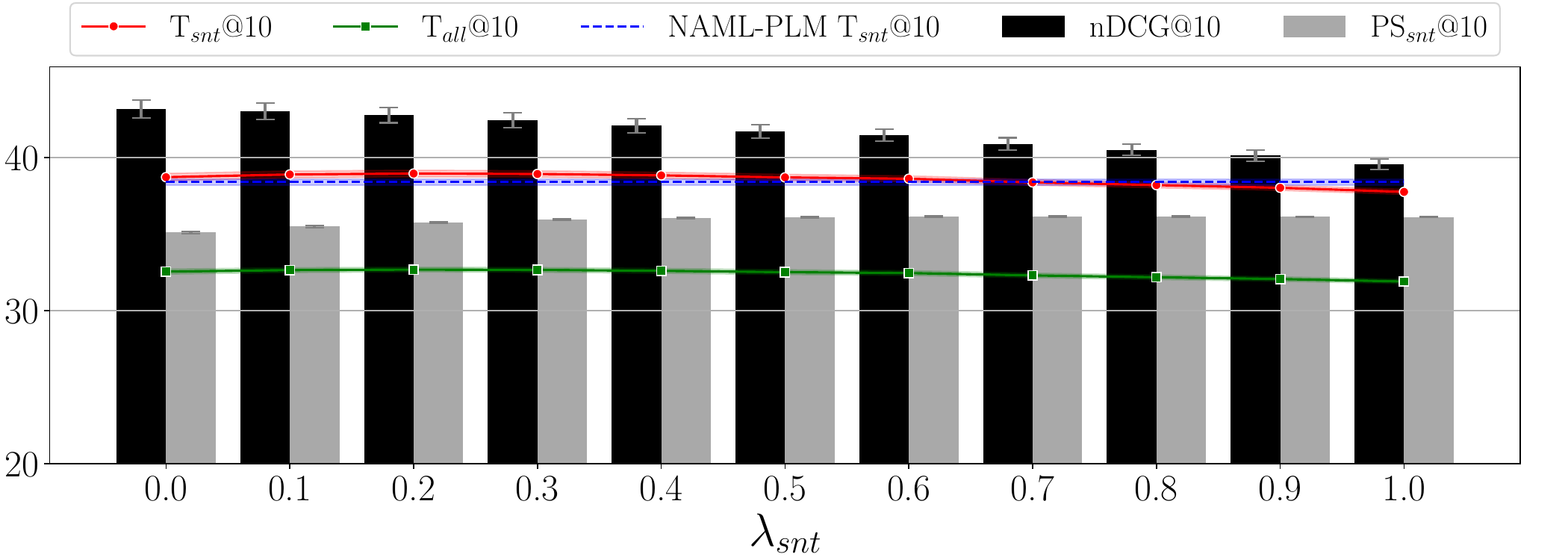}
         \caption{Single-aspect sentiment personalization.}
         \label{fig:single_aspect_pers_sent_mind}
     \end{subfigure}
     \hfill
     \begin{subfigure}[b]{0.48\textwidth}
         \centering
         \includegraphics[width=\textwidth]{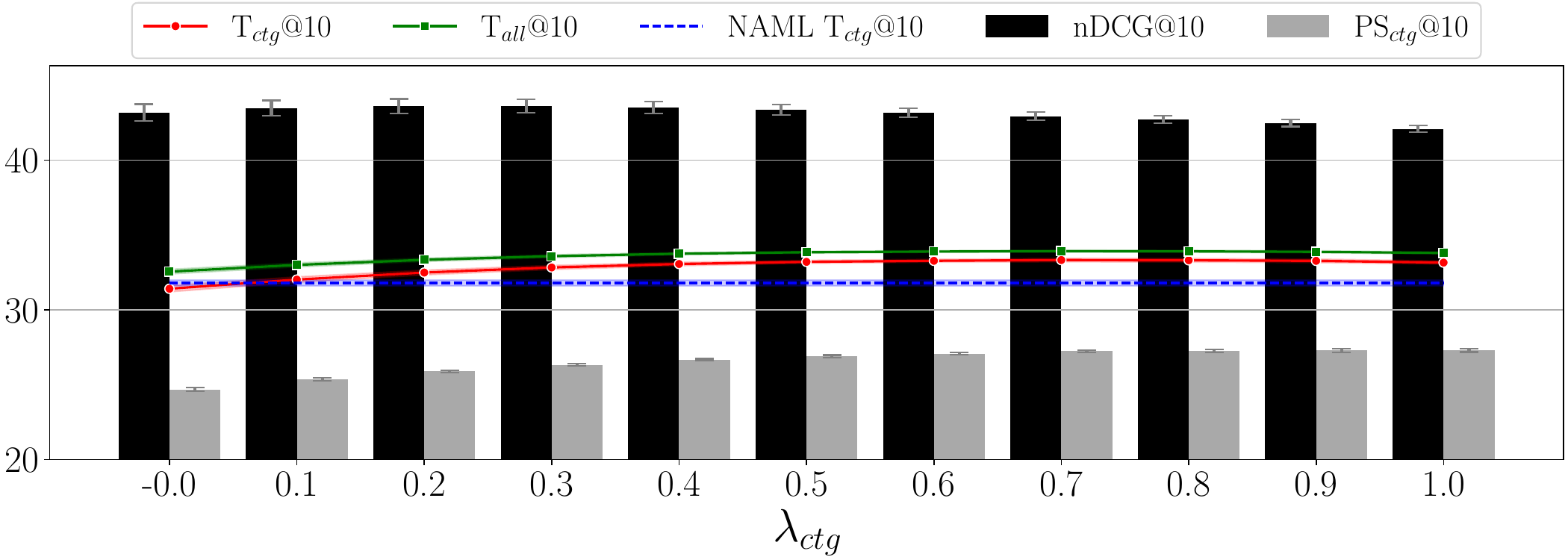}
         \caption{Single-aspect category personalization.}
         \label{fig:single_aspect_pers_categ_mind}
     \end{subfigure}
    \caption{Results of single-aspect customization for \manner{} and the best baseline on MIND. 
    }
    \label{fig:single_aspect_results_mind}
    \vspace{-1em}
\end{figure*}
\begin{figure*}[ht]
     \centering
    \begin{subfigure}[b]{0.48\textwidth}
         \centering
         \includegraphics[width=\textwidth]{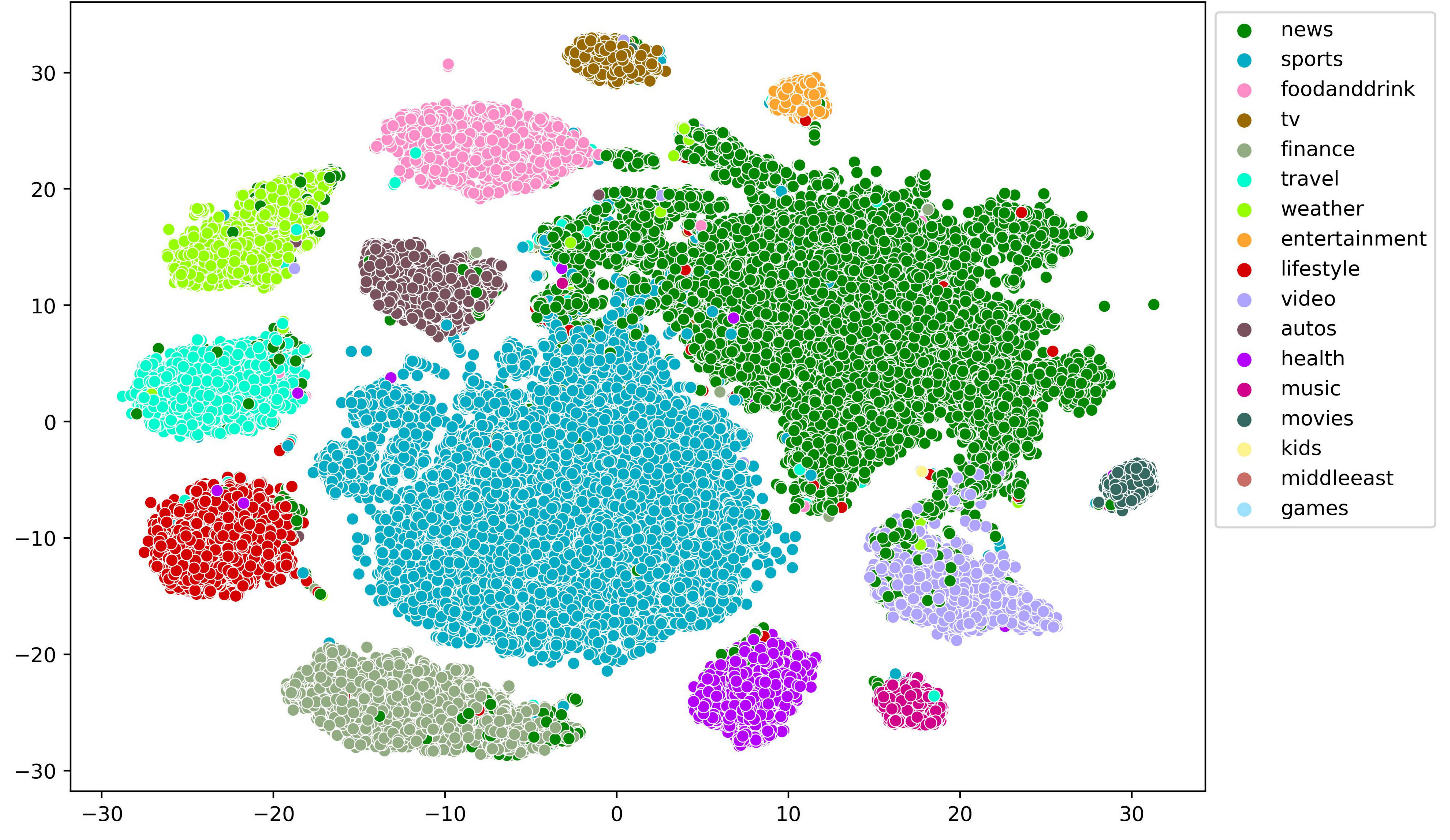}
         \caption{Category-shaped embedding space.}
         \label{fig:tsne_categ_mind}
     \end{subfigure}
     \hfill
     \begin{subfigure}[b]{0.48\textwidth}
         \centering
         \includegraphics[width=0.95\textwidth]{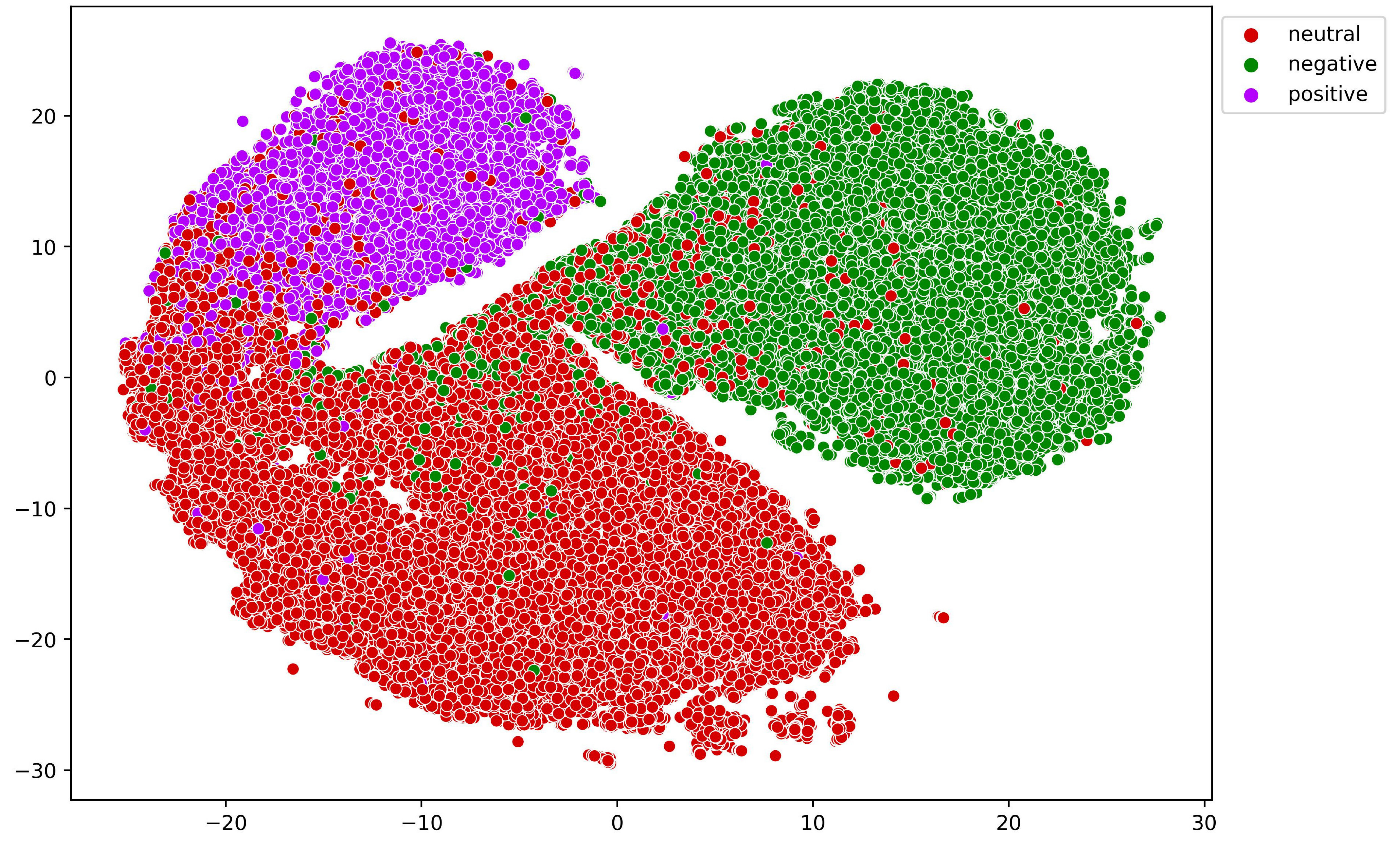}
         \caption{Sentiment-shaped embedding space.}
         \label{fig:tsne_sent_mind}
     \end{subfigure}
    \caption{t-SNE plots of the news embeddings in the test set of MIND.}
    \label{fig:tsne_embeddings_mind}
    \vspace{-0.5em}
\end{figure*}

\vspace{1.4mm}
\noindent\textbf{Diversification.}
Table~\ref{tab:results_diversification} summarizes the results on aspect diversification tasks.
Most baselines (including \manner{}'s \texttt{CR-Module} without aspect diversification) obtain similar diversification scores (D\textsubscript{ctg} and D\textsubscript{snt}). 
The sentiment-aware SentiRec-PLM, with an explicit auxiliary sentiment diversification objective, yields the highest sentiment diversity on Adressa; this comes at the expense of content personalization quality (lowest nDCG).
On MIND, the sentiment-specific SentiDebias-PLM achieves the highest sentiment diversity, but also exhibits lower content personalization performance. Overall, these results point to a trade-off between content personalization and aspectual diversity: models with higher D\textsubscript{A\textsubscript{p}} tend to have a lower nDCG. 

Unlike all other models, \manner{} can trade content personalization for diversity (and vice-versa) with different values of the aspect coefficients $\lambda_{A_p}$. Figs. \ref{fig:single_aspect_div_sent_mind}-\ref{fig:single_aspect_div_categ_mind} illustrate its performance in single-aspect sentiment and category diversification tasks for different values of $\lambda$\textsubscript{snt}, and $\lambda$\textsubscript{ctg}, respectively, on MIND.
The steady drop in nDCG together with the steady increase in D\textsubscript{A\textsubscript{p}} indeed indicate the existence of a trade-off between content personalization and aspect diversification. For topical categories we observe a steeper decline in content personalization quality with improved diversification than for sentiment. Sentiment diversity reaches peak performance for $\lambda_{snt} \hspace{-0.2em} = \hspace{-0.2em} -0.4$, whereas category diversity continues to increase up to $\lambda_{ctg} \hspace{-0.2em} = \hspace{-0.2em} -0.9$. Intuitively, content-based recommendation is more aligned with the topical than with the sentiment consistency of recommendations. The best trade-off (i.e., maximal performance w.r.t. T\textsubscript{A\textsubscript{p}}@10) is achieved for $\lambda_{snt} \hspace{-0.2em} = \hspace{-0.2em} -0.3$ for sentiment, and $\lambda_{ctg}\hspace{-0.2em} = \hspace{-0.2em} -0.2$ for topics.\footnote{We report analogous results on Adressa in Figs.  \ref{fig:single_aspect_div_sent_adressa}-\ref{fig:single_aspect_div_categ_adressa}.}
We attribute these effects to the representation spaces of the \texttt{A-Modules}. Fig. \ref{fig:tsne_embeddings_mind} shows the 2-dimensional t-SNE visualizations \cite{van2008visualizing} of the news embeddings produced with category-specialized, and respectively, sentiment-specialized NEs trained on MIND. The results confirm that the  encoder's latent representation space was reshaped to group same-class instances. The separation of classes, however, is less prominent for representation spaces of the encoders trained on Adressa (cf. Fig. \ref{fig:tsne_embeddings_adressa},  e.g., the effect is stronger on the category-shaped embedding space).\footnote{We believe that this is because Adressa has 10 times fewer news than MIND, with over half of the topical categories in Adressa being represented with fewer than 100 examples.} 
\newcolumntype{g}{>{\columncolor{Gray}}c}

\begin{table*}[ht]
\centering
\resizebox{\textwidth}{!}{%
    \begin{tabular}{l|cgcgcg|cgcgcg}
        \toprule
        \multicolumn{1}{c}{} & \multicolumn{6}{c}{\textbf{MIND}} & \multicolumn{6}{c}{\textbf{Adreesa}} \\  
        \cmidrule(lr){2-7} \cmidrule(lr){8-13}
        
        \multicolumn{1}{l|}{Model} 
        & nDCG@10 
        & PS\textsubscript{ctg}@10 & T\textsubscript{ctg}@10 
        & PS\textsubscript{snt}@10 & T\textsubscript{snt}@10 
        & T\textsubscript{all}@10
        
        & nDCG@10 
        & PS\textsubscript{ctg}@10 & T\textsubscript{ctg}@10 
        & PS\textsubscript{snt}@10 & T\textsubscript{snt}@10 
        & T\textsubscript{all}@10
        \\
        \hline

       NRMS-PLM 
        & 39.9\textsubscript{$\pm$0.6} 
        & 23.9\textsubscript{$\pm$0.2} 
        & 29.9\textsubscript{$\pm$0.3} 
        & 35.1\textsubscript{$\pm$0.1} 
        & 37.3\textsubscript{$\pm$0.3} 
        & 31.5\textsubscript{$\pm$0.2} 

        & 51.3\textsubscript{$\pm$2.3} 
        & 34.3\textsubscript{$\pm$0.4} 
        & 41.1\textsubscript{$\pm$1.0} 
        & 41.8\textsubscript{$\pm$0.1} 
        & 46.1\textsubscript{$\pm$1.0} 
        & 41.3\textsubscript{$\pm$0.7} 
        \\
        
        MINER
        & 40.0\textsubscript{$\pm$1.0} 
        & 23.9\textsubscript{$\pm$0.4} 
        & 29.9\textsubscript{$\pm$0.5} 
        & 35.0\textsubscript{$\pm$0.2} 
        & 37.3\textsubscript{$\pm$0.4} 
        & 31.5\textsubscript{$\pm$0.4} 

        & 46.3\textsubscript{$\pm$4.1} 
        & 34.4\textsubscript{$\pm$0.2} 
        & 39.4\textsubscript{$\pm$1.5} 
        & 42.0\textsubscript{$\pm$0.0} 
        & 43.9\textsubscript{$\pm$1.8} 
        & 40.2\textsubscript{$\pm$1.0} 
        \\
        \hdashline

        NAML-PLM
        & 42.2\textsubscript{$\pm$0.4} 
        & \underline{25.5\textsubscript{$\pm$0.2}} 
        & \underline{31.8\textsubscript{$\pm$0.2}} 
        & 35.1\textsubscript{$\pm$0.2} 
        & 38.4\textsubscript{$\pm$0.2} 
        & \underline{32.8\textsubscript{$\pm$0.2}} 

        & 52.5\textsubscript{$\pm$4.1} 
        & \underline{36.1\textsubscript{$\pm$0.8}} 
        & \underline{42.7\textsubscript{$\pm$1.7}} 
        & 41.8\textsubscript{$\pm$0.1} 
        & 46.5\textsubscript{$\pm$1.7} 
        & \underline{42.4\textsubscript{$\pm$1.1}} 
        \\
        
        LSTUR-PLM
        & 38.3\textsubscript{$\pm$1.7} 
        & 24.0\textsubscript{$\pm$1.0} 
        & 29.5\textsubscript{$\pm$1.2} 
        & 34.8\textsubscript{$\pm$0.3} 
        & 36.5\textsubscript{$\pm$0.9} 
        & 31.1\textsubscript{$\pm$1.0} 

        & 51.2\textsubscript{$\pm$2.0} 
        & 35.1\textsubscript{$\pm$2.1} 
        & 41.6\textsubscript{$\pm$1.0} 
        & 41.8\textsubscript{$\pm$0.1} 
        & 46.0\textsubscript{$\pm$0.8} 
        & 41.7\textsubscript{$\pm$0.7} 
        \\

        MINS-PLM
        & 40.8\textsubscript{$\pm$0.3} 
        & 25.0\textsubscript{$\pm$0.3} 
        & 31.0\textsubscript{$\pm$0.3} 
        & 34.7\textsubscript{$\pm$0.2} 
        & 37.5\textsubscript{$\pm$0.2} 
        & 32.1\textsubscript{$\pm$0.3} 

        & 53.0\textsubscript{$\pm$3.4} 
        & 33.9\textsubscript{$\pm$0.7} 
        & 41.3\textsubscript{$\pm$1.4} 
        & 41.8\textsubscript{$\pm$0.1} 
        & 46.7\textsubscript{$\pm$1.3} 
        & 41.5\textsubscript{$\pm$1.0} 
        \\

        CAUM\textsubscript{no entities}-PLM
        & 41.0\textsubscript{$\pm$1.9} 
        & 24.8\textsubscript{$\pm$0.6} 
        & 30.9\textsubscript{$\pm$1.0} 
        & 35.0\textsubscript{$\pm$0.2} 
        & 37.8\textsubscript{$\pm$0.9} 
        & 32.2\textsubscript{$\pm$0.7} 

        & 52.0\textsubscript{$\pm$1.2} 
        & 33.5\textsubscript{$\pm$0.2} 
        & 39.6\textsubscript{$\pm$1.1} 
        & 40.8\textsubscript{$\pm$0.4} 
        & 46.3\textsubscript{$\pm$0.5} 
        & 41.1\textsubscript{$\pm$0.3} 
        \\

        CAUM-PLM 
        & 40.8\textsubscript{$\pm$1.3} 
        & 25.1\textsubscript{$\pm$0.3} 
        & 31.1\textsubscript{$\pm$0.4} 
        & 35.0\textsubscript{$\pm$0.1} 
        & 37.7\textsubscript{$\pm$0.6} 
        & 32.3\textsubscript{$\pm$0.3} 

        & -- 
        & -- 
        & -- 
        & -- 
        & -- 
        & -- 
        \\

        TANR-PLM 
        & 41.6\textsubscript{$\pm$0.9} 
        & 25.2\textsubscript{$\pm$0.5} 
        & 31.4\textsubscript{$\pm$0.6} 
        & 35.0\textsubscript{$\pm$0.2} 
        & 38.0\textsubscript{$\pm$0.4} 
        & 32.5\textsubscript{$\pm$0.5} 

        & 51.4\textsubscript{$\pm$0.6} 
        & 34.0\textsubscript{$\pm$0.5} 
        & 41.0\textsubscript{$\pm$0.5} 
        & 41.8\textsubscript{$\pm$0.1} 
        & 46.1\textsubscript{$\pm$0.3} 
        & 41.2\textsubscript{$\pm$0.4} 
        \\
        \hdashline

        SentiRec-PLM
        & 40.5\textsubscript{$\pm$0.4} 
        & 24.2\textsubscript{$\pm$0.3} 
        & 30.3\textsubscript{$\pm$0.3} 
        & 34.6\textsubscript{$\pm$0.0} 
        & 37.3\textsubscript{$\pm$0.2} 
        & 31.6\textsubscript{$\pm$0.2} 

        & 40.8\textsubscript{$\pm$2.4} 
        & 32.4\textsubscript{$\pm$0.3} 
        & 36.1\textsubscript{$\pm$1.0} 
        & 39.3\textsubscript{$\pm$0.1} 
        & 40.0\textsubscript{$\pm$1.2} 
        & 37.1\textsubscript{$\pm$0.7} 
        \\ 
        
        SentiDebias-PLM
        & 32.2\textsubscript{$\pm$2.0} 
        & 20.8\textsubscript{$\pm$1.3} 
        & 25.2\textsubscript{$\pm$1.5} 
        & 34.1\textsubscript{$\pm$0.3} 
        & 33.1\textsubscript{$\pm$1.2} 
        & 27.6\textsubscript{$\pm$1.2} 

        & 44.2\textsubscript{$\pm$2.9} 
        & 34.1\textsubscript{$\pm$0.6} 
        & 38.5\textsubscript{$\pm$1.3} 
        & 41.8\textsubscript{$\pm$0.1} 
        & 42.9\textsubscript{$\pm$1.4} 
        & 39.5\textsubscript{$\pm$1.0} 
        \\
        \hline

        \manner{} (\texttt{CR-Module})
        & \textbf{43.2\textsubscript{$\pm$0.6}}
        & 24.7\textsubscript{$\pm$0.1} 
        & 31.4\textsubscript{$\pm$0.2} 
        & 35.1\textsubscript{$\pm$0.1} 
        & 38.7\textsubscript{$\pm$0.2} 
        & 32.6\textsubscript{$\pm$0.2} 

        & \textbf{54.3\textsubscript{$\pm$2.5}} 
        & 34.5\textsubscript{$\pm$0.1} 
        & 42.2\textsubscript{$\pm$0.8} 
        & 42.0\textsubscript{$\pm$0.1} 
        & \underline{47.3\textsubscript{$\pm$0.9}} 
        & 42.1\textsubscript{$\pm$0.5} 
        \\

        \manner{} ($\lambda_{ctg}=0.7 / 0.4$, $\lambda_{snt}=0$)
        & \underline{42.9\textsubscript{$\pm$0.3}} 
        & \textbf{27.2\textsubscript{$\pm$0.1}} 
        & \textbf{33.3\textsubscript{$\pm$0.1}} 
        & \underline{35.2\textsubscript{$\pm$0.0}} 
        & \underline{38.7\textsubscript{$\pm$0.1}} 
        & \textbf{33.9\textsubscript{$\pm$0.1}} 

         & 53.6\textsubscript{$\pm$1.9} 
        & \textbf{36.2\textsubscript{$\pm$0.1}} 
        & \textbf{43.2\textsubscript{$\pm$0.7}} 
        & \underline{42.1\textsubscript{$\pm$0.1}} 
        & 47.2\textsubscript{$\pm$0.7} 
        & \textbf{42.9\textsubscript{$\pm$0.4}} 
        \\

        \manner{} ($\lambda_{ctg}=0$, $\lambda_{snt}=0.2 / 0.1$)
        & 42.8\textsubscript{$\pm$0.5} 
        & 24.7\textsubscript{$\pm$0.1} 
        & 31.3\textsubscript{$\pm$0.2} 
        & \textbf{35.8\textsubscript{$\pm$0.1}} 
        & \textbf{39.0\textsubscript{$\pm$0.2}} 
        & 32.7\textsubscript{$\pm$0.1} 

       & \underline{54.1\textsubscript{$\pm$2.4}} 
        & 34.7\textsubscript{$\pm$0.1} 
        & 42.2\textsubscript{$\pm$0.8} 
        & \textbf{42.2\textsubscript{$\pm$0.1}} 
        & \textbf{47.4\textsubscript{$\pm$0.9}} 
        & 42.2\textsubscript{$\pm$0.5} 
        \\ 

        \bottomrule
        
    \end{tabular}%
    }
\caption{Single-aspect personalization. For \manner{}, we list the best results (cf. T\textsubscript{A\textsubscript{p}}) of single-aspect diversification as $\lambda_{A_p}$ (MIND/Adressa). The best results per column are highlighted in bold, the second best underlined.}
\label{tab:results_personalization}

\vspace{-0.5em}
\end{table*}

\vspace{1.4mm}
\noindent\textbf{Personalization.}
Table \ref{tab:results_personalization} displays the results on aspect personalization tasks. TANR, trained with an auxiliary topic classification task, underperforms NAML, which uses topical categories as NE input features, in category personalization on both datasets. 
\manner{}'s \texttt{CR-Module} alone (i.e., without any aspect customization) yields competitive category personalization performance. We believe that this is because (i) the \texttt{CR-Module} is best in content personalization and (ii) category personalization is well-aligned with content personalization (i.e., news with similar content tend to belong to the same category). 
Fig. \ref{fig:single_aspect_pers_categ_mind} explores the trade-off between content and category personalization, for positive values of $\lambda_{ctg}$ on MIND.
%
The best topical category personalization (PS\textsubscript{ctg}), obtained for $\lambda_{ctg} \hspace{-0.2em} > \hspace{-0.2em} 0.7$, comes at the small expense of content personalization: too much weight on the category similarity of news dilutes the impact of content relevance. Increased sentiment personalization (cf. Fig. \ref{fig:single_aspect_pers_sent_mind}), however, is much more detrimental to content personalization. Intuitively, users do not choose articles based on sentiment. Tailoring recommendations according to the sentiment of previously clicked news thus leads to more content-irrelevant suggestions.

\subsection{Multi-Aspect Customization} 
\begin{figure*}[ht]
     \centering
     \begin{subfigure}[b]{0.48\textwidth}
         \centering
         \includegraphics[width=\textwidth]{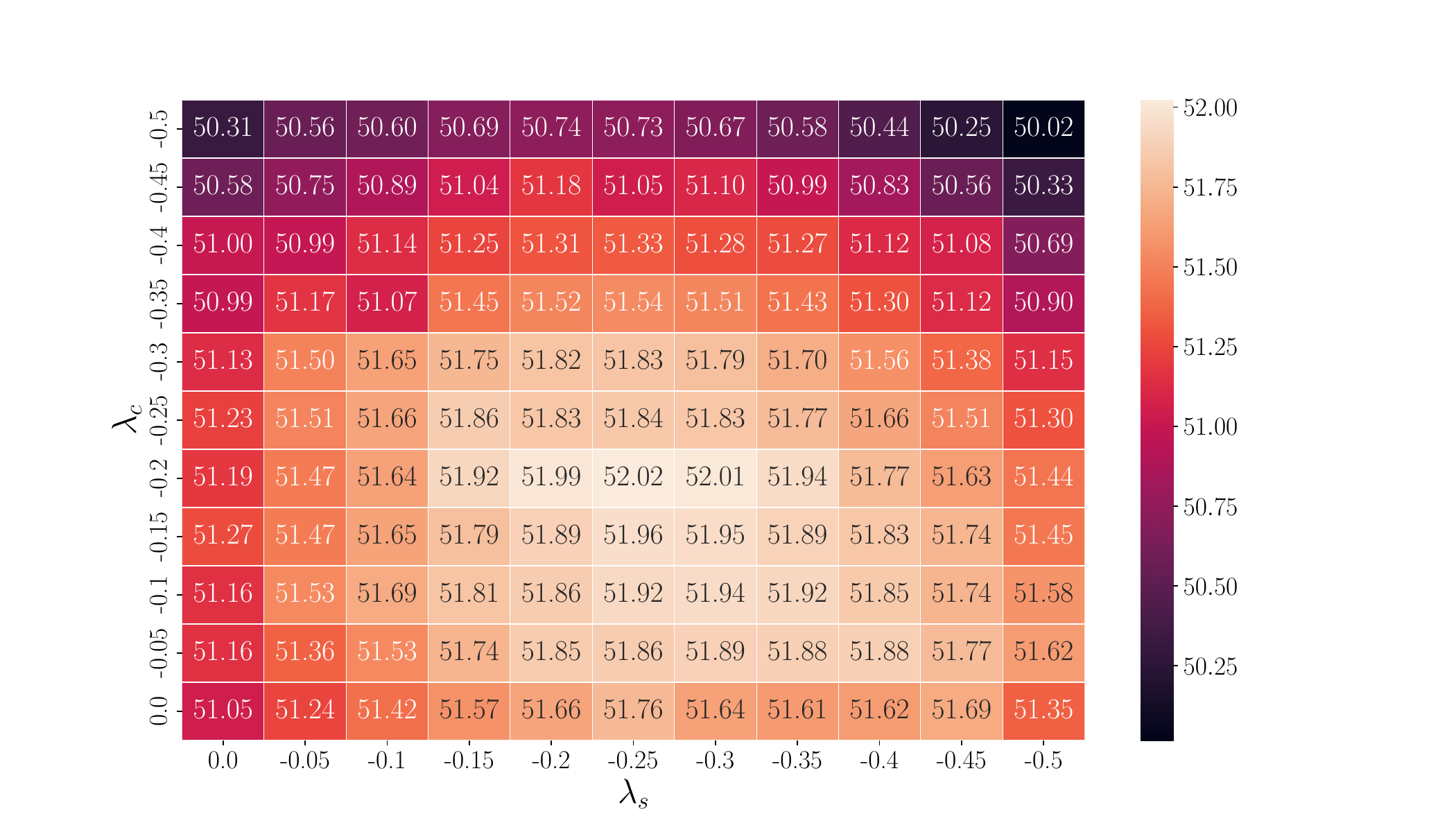}
         \caption{Multi-aspect diversification on MIND.}
         \label{fig:multi_aspect_div_mind}
     \end{subfigure}
     \hfill
     \begin{subfigure}[b]{0.48\textwidth}
         \centering
         \includegraphics[width=\textwidth]{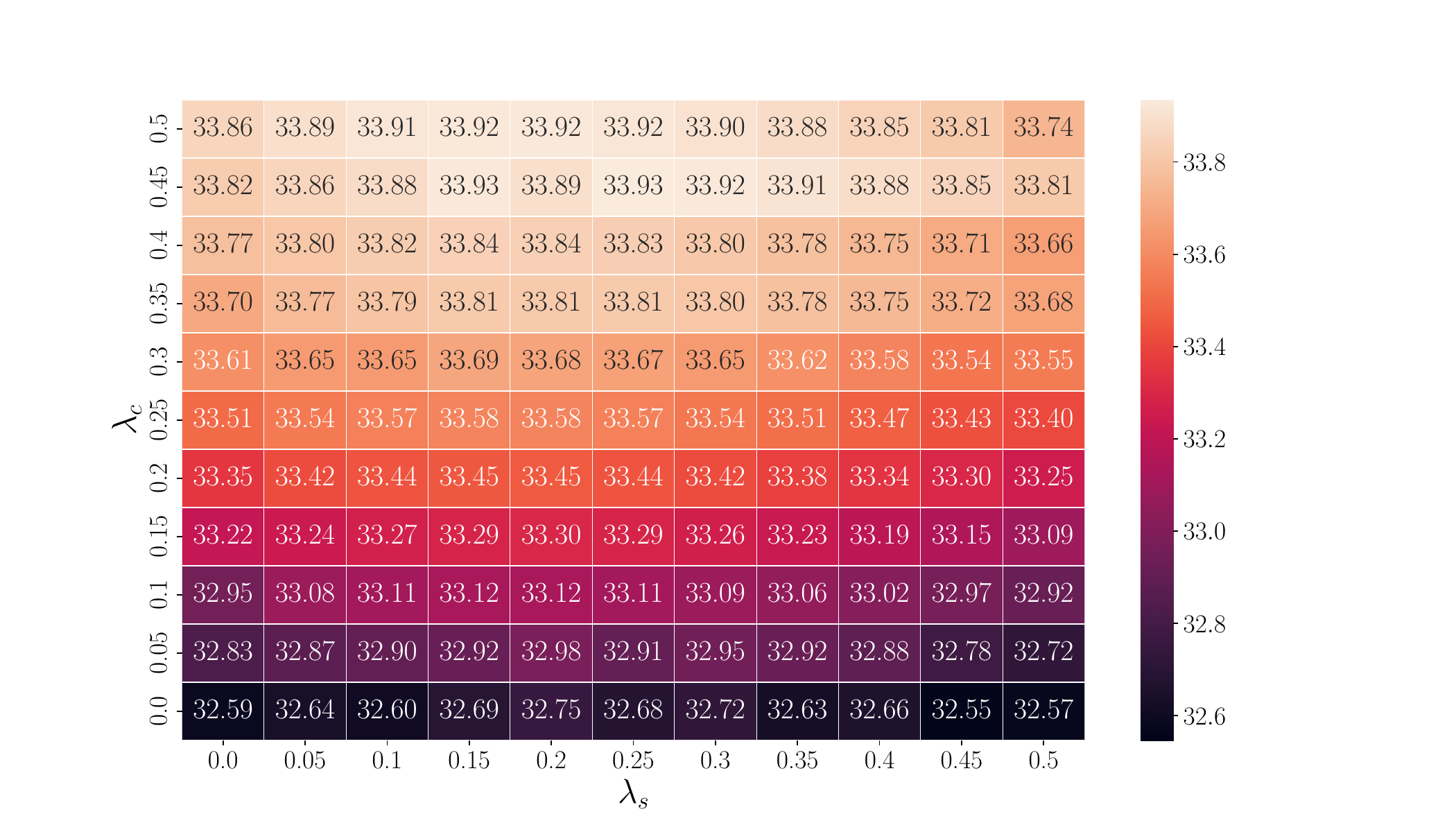}
         \caption{Multi-aspect personalization on MIND.}
         \label{fig:multi_aspect_pers_mind}
     \end{subfigure}
    \caption{Results of multi-aspect customization for \manner{} on MIND.}
    \label{fig:multi_aspect_results_mind}
    \vspace{-0.5em}
\end{figure*}

We further explore the trade-off between content personalization and multi-aspect diversification, i.e. diversifying over both topical categories and sentiments.
We achieve the highest T\textsubscript{all} for $\lambda_{ctg} \hspace{-0.2em} = \hspace{-0.2em} -0.2$ and $\lambda_{snt} \hspace{-0.2em} = \hspace{-0.2em} -0.25$ on MIND (cf. Fig. \ref{fig:multi_aspect_div_mind}). In line with single-aspect diversification results, we observe that improving diversity in terms of topical categories rather than sentiments has a more negative effect on content personalization quality, i.e. steeper decline in T\textsubscript{all}. These results confirm that \manner{} can generalize to diversify for multiple aspects at once by weighting individual aspect relevance scores less than in the single-aspect task. Weighting several aspects higher simultaneously acts as a double discounting for content personalization, diluting content relevance disproportionately.
Similarly, for multi-aspect personalization, we achieve the best multi-aspect trade-off on MIND (cf. Fig. \ref{fig:multi_aspect_pers_mind}) for $\lambda_{ctg} \hspace{-0.2em} = \hspace{-0.2em} 0.45$ and $\lambda_{snt} \hspace{-0.2em} = \hspace{-0.2em} 0.25$. Stronger enforcing of alignment of candidate news with the user's history is needed for topical categories than for sentiment (i.e., $\lambda_{ctg} \hspace{-0.2em} > \hspace{-0.2em} \lambda_{snt}$). This is because sentiment exhibits low variance within categories (e.g., \textit{politics} news are mostly negative) and enforcing categorical personalization partly also achieves sentiment personalization.\footnote{We refer to Fig. \ref{fig:multi_aspect_results_addressa} for analogous results on Adressa.}

\begin{figure*}[ht]
     \centering
     \begin{subfigure}[b]{0.48\textwidth}
         \centering
         \includegraphics[width=\textwidth]{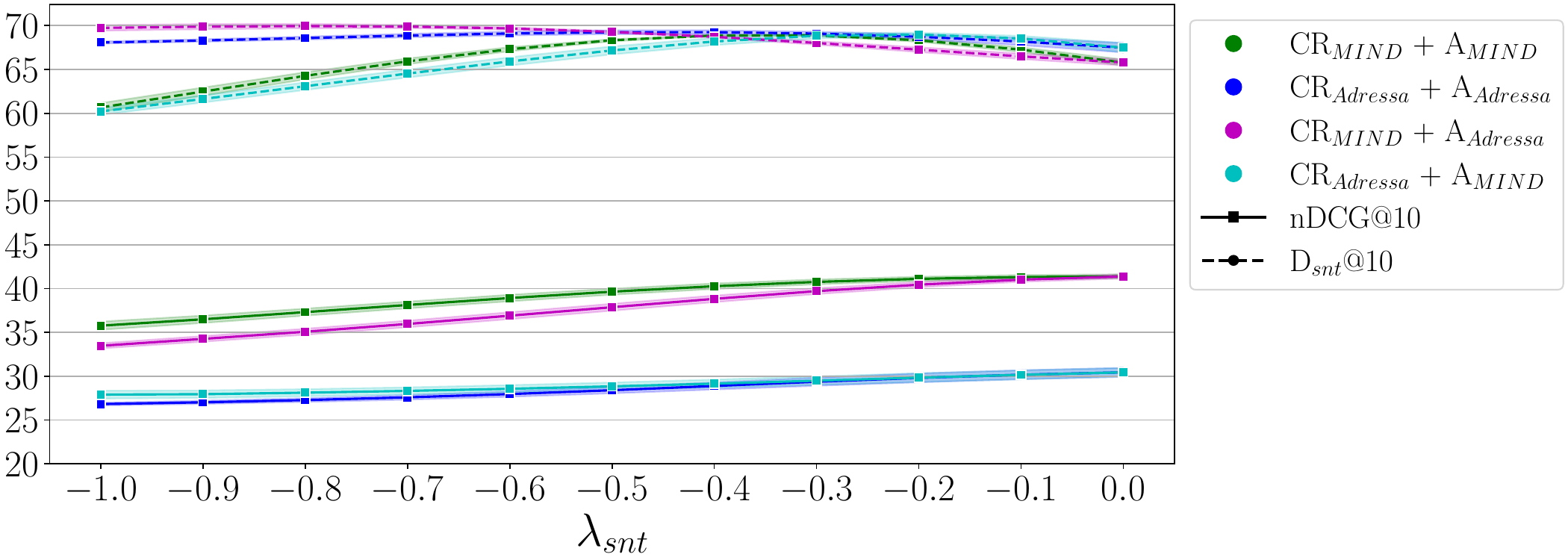}
         \caption{Sentiment diversification.}
         \label{fig:adressa_transfer_mind_div_sent}
     \end{subfigure}
     \hfill
     \begin{subfigure}[b]{0.48\textwidth}
         \centering
         \includegraphics[width=\textwidth]{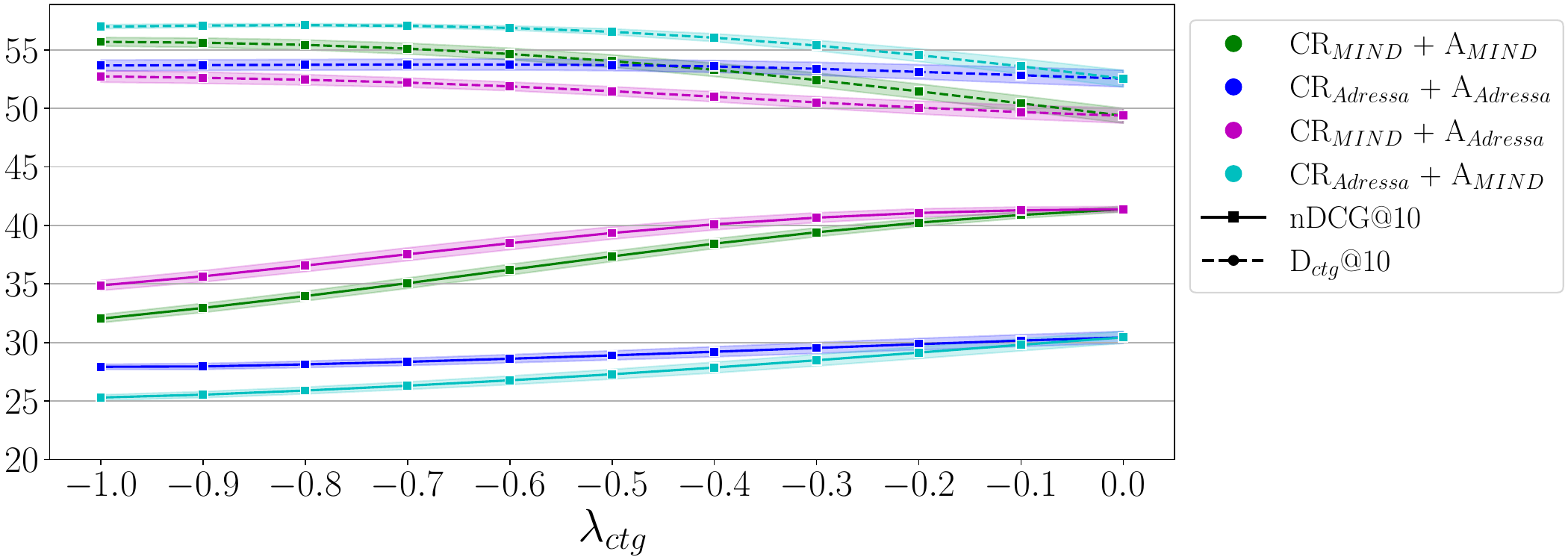}
         \caption{Category diversification.}
         \label{fig:adressa_transfer_mind_div_categ}
     \end{subfigure}
    \caption{\texttt{XLT} in single-aspect diversification, with modules trained on different (combinations of) source-language datasets and evaluated on the target-language dataset MIND. The line style indicates the metric, the color the source-language datasets used in training. 
    }
    \label{fig:xlt_adressa_mind_div}
    \vspace{-1em}
\end{figure*}

\subsection{Cross-Lingual Transfer}
\label{sec:xlt}

We next analyze the transferability of \manner{} across datasets and languages in single-aspect customization experiments.\footnote{We evaluate only the title-based version of \manner{}, as the full version cannot be trained on Adressa.} 
Concretely, we train the \texttt{CR-Module} and \texttt{A-Modules} on both MIND (i.e., in English) and Adressa (i.e., in Norwegian), respectively. 
At inference, we evaluate all combinations of pretrained \texttt{CR-Module} and \texttt{A-Modules} on the test set of MIND. 
We now use a multilingual DistilBERT Base \cite{sanh2019distilbert} as \manner{}'s NE to enable cross-lingual transfer (\texttt{XLT}).
Fig. \ref{fig:xlt_adressa_mind_div} summarizes the \texttt{XLT} results for single-aspect diversification.\footnote{Figs. \ref{fig:xlt_adressa_mind_div} and \ref{fig:xlt_mind_adressa} provide similar results for single-aspect personalization and single-aspect customization on MIND, and respectively, Adressa, as target-language datasets.}
As expected, \manner{} trained fully on Adressa suffers a large drop in content personalization performance, compared to the counterpart trained on MIND. 
In contrast, transferring only the \texttt{A-Module}, i.e., training the \texttt{CR-Module} on MIND and the \texttt{A-Module} on Adressa, yields performance comparable to that of complete in-language training (i.e., both \texttt{CR-Module} and \texttt{A-Module} trained on MIND). 
This is particularly the case for the sentiment \texttt{A-Module}, since the sentiment labels between the datasets are more aligned than those for topical categories. 
These results indicate that the plug-and-play of \texttt{A-Modules} enables zero-shot \texttt{XLT}, as modules trained on the much smaller Norwegian Adressa transfer well to the large English MIND. 
This suggests that, coupled with multilingual PLMs, \manner{} can be used for effective news recommendation in lower-resource languages, where training data and aspectual labels are scarce. Furthermore, the results demonstrate that the \texttt{A-Modules} could be trained on general-purpose classification datasets  (e.g. topic or sentiment classification datasets), 
alleviating the need for aspect-specific annotation of news stories. 

\subsection{Computational Complexity}
\label{sec:computational_complexity}
While A-Modules add extra parameters, their average training time is two orders of magnitude faster than that of the CR-Module.\footnote{On MIND (Adressa), the A-Module for topical category trains 277 (51) times faster and that for sentiment 204 (53) times faster on average per epoch than the CR-Module.}
This is a \textit{one-time} increase in training time: the resulting modules can then be arbitrarily combined for any recommendation goal without additional training. In contrast, all other NNRs require re-training or fine-tuning if the recommendation objective changes as the model weights have to be adjusted each time. This translates into much higher computational costs in practice. We emphasize that \manner{} also has a much lower inference latency due to the (i) CR-Module's lean architecture without a parameterized UE, and (ii) ability to parallelize loading and deploying different modules, for which only the final score has to be combined.\footnote{We provide the average inference times in Appendix \ref{sec:appendix_time}.} 
Overall, considering both training and inference, \manner{} is more efficient and flexible in a realistic setup with differing recommendation goals that may vary by user or for the same user over time. 
\section{Conclusion}
\label{sec:conclusion}

We proposed \manner{}, a modular framework for multi-aspect neural news recommendation. It learns aspect-specialized NEs with supervised contrastive learning, and linearly combines the corresponding aspect-specific similarity scores for final ranking. Its modular design allows defining ad-hoc multi-aspect ranking functions at inference. 
Our experiments show that \manner{} consistently outperforms state-of-the-art NNRs on both (i) standard content-based recommendation, and on single- and multi-aspect (ii) diversification and (iii) personalization of recommendations. Moreover, we can identify on-the-fly optimal trade-offs between content-based recommendation performance and aspect-based customization.
Equipped with a multilingual PLM, \manner{} can successfully cross-lingually transfer aspect-specific NEs. This supports use cases where aspect-specific labels (e.g., sentiment) are not available for news in the target languages of interest. We hope that our work stimulates more research towards modular, easily extendable, and reusable news recommenders.   

\section*{Limitations}
\label{sec:limitations}

\manner{} targets exclusively content-based neural news recommendation and leverages solely textual features. In practice, recommender systems may incorporate content features from various other modalities (e.g., image, video), as well as similarities between users in a collaborative filtering manner. We leave the extension of \manner{} with multi-modal content for future work.

\manner{} independently handles each aspect and aggregates them by weighting the aspect-specific similarities. While it could be argued that direct interactions between different aspects might improve the recommendation performance, training a separate A-Module for each aspect is exactly what drives \manner's flexibility. The A-modules allow the user to arbitrarily define the preferences for any concrete recommendation, by defining how much diversification or personalization is desired over each aspect. As illustrated by the results of our experiments, \manner{} outperforms all the state-of-the-art systems, including the ones where additional aspects are directly integrated in the training objective or in the news encoder. MANNeR is thus, besides being drastically more flexible (as it supports arbitrary recommendation objectives at inference time), also more performant, despite the fact that no interactions exist between the aspect modules at training time.

Our framework fully fine-tunes a PLM per aspect-specific module (either for content-relevance in the \texttt{CR-Module} or for aspect similarity in the \texttt{A-Module}). As all modules share the same PLM as backbone, parameter efficient fine-tuning (PEFT), e.g.  LoRA \cite{hu2021lora}, would bypass the need to repeatedly load the entire PLM per module into memory. PEFT has been shown to closely meet the performance of full fine-tuning. This represents a key advantage for deploying \manner{} in real-world applications. We however fully fine-tuned models to avoid PEFT as a confounding factor in our experiments. We further chose base-sized PLMs as the backbone of the NE in all models due to computational constraints. While in fine-tuning they remain competitive to large language models (LLMs), the latter may capture richer semantics, which can prove particularly useful for \texttt{XLT} applications. With a PEFT approach, \manner{} could easily leverage LLMs without a corresponding increase in computational resources.

Lastly, there exist varied approaches for measuring the descriptive \cite{castells2021novelty} and normative \cite{vrijenhoek2023radio} diversity of recommendations. While some of these metrics can be tailored to support arbitrary aspects (i.e., to measure the diversity of recommendations w.r.t. to a particular categorical feature), we opted to quantify aspect-based diversity as generally as possible, leveraging only the distribution of an aspect's values in the recommendation list. We leave exploration of further diversity metrics to future work.

\section*{Ethical Considerations}
\label{sec:ethical_consideations}

We consider several ethical considerations that arise when working with recommender systems and  open benchmark datasets. On the one hand, any biases or misinformation that might exist in the news and user data provided in the public datasets could be propagated through the recommendation pipeline. Similarly, the PLMs used as the recommenders' backbone could contain social biases captured from the training data. On the other hand, the \texttt{A-Modules} in \manner{} could be abused to reduce the diversity of recommendations by over-weighting the aspectual-similarity with the user's history, particularly for sensitive aspects such as news stance. This, in turn, could lead to reinforcing the users' existing worldviews and stances \cite{li2019survey}. Therefore, safeguards should be incorporated in the recommendation models to ensure not only that the outputs are accurate and truthful, but also that the systems are not misused to constrain access to diverse viewpoints. 
\section*{Acknowledgments}

The authors acknowledge support by the state of Baden-Württemberg through bwHPC and the German Research Foundation (DFG) through grant INST 35/1597-1 FUGG. Andreea Iana was supported by the ReNewRS project grant, which is  funded by the Baden-Württemberg Stiftung in the Responsible Artificial Intelligence program.

\bibliography{custom}

\begin{thebibliography}{69}
\providecommand{\natexlab}[1]{#1}

\bibitem[{An et~al.(2019)An, Wu, Wu, Zhang, Liu, and Xie}]{an2019lstur}
Mingxiao An, Fangzhao Wu, Chuhan Wu, Kun Zhang, Zheng Liu, and Xing Xie. 2019.
\newblock \href {https://doi.org/10.18653/v1/P19-1033} {Neural news recommendation with long-and short-term user representations}.
\newblock In \emph{Proceedings of the 57th Annual Meeting of the Association for Computational Linguistics}, pages 336--345.

\bibitem[{Bahdanau et~al.(2014)Bahdanau, Cho, and Bengio}]{bahdanau2014neural}
Dzmitry Bahdanau, Kyunghyun Cho, and Yoshua Bengio. 2014.
\newblock \href {https://arxiv.org/pdf/1409.0473.pdf} {Neural machine translation by jointly learning to align and translate}.
\newblock \emph{ICLR}.

\bibitem[{Barbieri et~al.(2022)Barbieri, Anke, and Camacho-Collados}]{barbieri2022xlm}
Francesco Barbieri, Luis~Espinosa Anke, and Jose Camacho-Collados. 2022.
\newblock \href {https://aclanthology.org/2022.lrec-1.27} {{XLM-T}: Multilingual language models in twitter for sentiment analysis and beyond}.
\newblock In \emph{Proceedings of the Thirteenth Language Resources and Evaluation Conference}, pages 258--266.

\bibitem[{Bonnici(2020)}]{bonnici2020kullback}
Vincenzo Bonnici. 2020.
\newblock \href {https://doi.org/10.48550/arXiv.2008.05932} {Kullback-leibler divergence between quantum distributions, and its upper-bound}.
\newblock \emph{arXiv preprint arXiv:2008.05932}.

\bibitem[{Bordes et~al.(2013)Bordes, Usunier, Garcia-Dur{\'a}n, Weston, and Yakhnenko}]{bordes2013translating}
Antoine Bordes, Nicolas Usunier, Alberto Garcia-Dur{\'a}n, Jason Weston, and Oksana Yakhnenko. 2013.
\newblock \href {https://dl.acm.org/doi/abs/10.5555/2999792.2999923} {Translating embeddings for modeling multi-relational data}.
\newblock In \emph{Proceedings of the 26th International Conference on Neural Information Processing Systems-Volume 2}, pages 2787--2795.

\bibitem[{Castells et~al.(2021)Castells, Hurley, and Vargas}]{castells2021novelty}
Pablo Castells, Neil Hurley, and Saul Vargas. 2021.
\newblock \href {https://doi.org/10.1007/978-1-0716-2197-4_16} {Novelty and diversity in recommender systems}.
\newblock In \emph{Recommender systems handbook}, pages 603--646. Springer.

\bibitem[{Chen et~al.(2017)Chen, Meng, Xu, and Lukasiewicz}]{chen2017location}
Cheng Chen, Xiangwu Meng, Zhenghua Xu, and Thomas Lukasiewicz. 2017.
\newblock \href {https://doi.org/10.1109/ACCESS.2017.2655150} {Location-aware personalized news recommendation with deep semantic analysis}.
\newblock \emph{IEEE Access}, 5:1624--1638.

\bibitem[{Choi et~al.(2022)Choi, Kim, and Gim}]{choi2022not}
Seonghwan Choi, Hyeondey Kim, and Manjun Gim. 2022.
\newblock \href {https://doi.org/10.1145/3487553.3524936} {Do not read the same news! enhancing diversity and personalization of news recommendation}.
\newblock In \emph{Companion Proceedings of the Web Conference 2022}, pages 1211--1215.

\bibitem[{Conneau et~al.(2020)Conneau, Khandelwal, Goyal, Chaudhary, Wenzek, Guzm{\'a}n, Grave, Ott, Zettlemoyer, and Stoyanov}]{conneau2020unsupervised}
Alexis Conneau, Kartikay Khandelwal, Naman Goyal, Vishrav Chaudhary, Guillaume Wenzek, Francisco Guzm{\'a}n, {\'E}douard Grave, Myle Ott, Luke Zettlemoyer, and Veselin Stoyanov. 2020.
\newblock \href {https://doi.org/10.18653/v1/2020.acl-main.747} {Unsupervised cross-lingual representation learning at scale}.
\newblock In \emph{Proceedings of the 58th Annual Meeting of the Association for Computational Linguistics}, pages 8440--8451.

\bibitem[{Freedman and Sears(1965)}]{freedman1965selective}
Jonathan~L Freedman and David~O Sears. 1965.
\newblock Selective exposure.
\newblock In \emph{Advances in experimental social psychology}, volume~2, pages 57--97. Elsevier.

\bibitem[{Gabriel De~Souza et~al.(2019)Gabriel De~Souza, Jannach, and Da~Cunha}]{gabriel2019contextual}
P~Moreira Gabriel De~Souza, Dietmar Jannach, and Adilson~Marques Da~Cunha. 2019.
\newblock \href {https://doi.org/10.1109/ACCESS.2019.2954957} {Contextual hybrid session-based news recommendation with recurrent neural networks}.
\newblock \emph{IEEE Access}, 7:169185--169203.

\bibitem[{Gao et~al.(2018)Gao, Xin, Liu, Wang, Lu, Li, Fan, and Guo}]{gao2018fine}
Jie Gao, Xin Xin, Junshuai Liu, Rui Wang, Jing Lu, Biao Li, Xin Fan, and Ping Guo. 2018.
\newblock \href {https://doi.org/10.1109/WI.2018.0-104} {Fine-grained deep knowledge-aware network for news recommendation with self-attention}.
\newblock In \emph{2018 IEEE/WIC/ACM International Conference on Web Intelligence (WI)}, pages 81--88. IEEE.

\bibitem[{Gharahighehi and Vens(2023)}]{gharahighehi2023diversification}
Alireza Gharahighehi and Celine Vens. 2023.
\newblock \href {https://doi.org/10.1007/s00779-021-01606-4} {Diversification in session-based news recommender systems}.
\newblock \emph{Personal and Ubiquitous Computing}, 27(1):5--15.

\bibitem[{Gulla et~al.(2017)Gulla, Zhang, Liu, {\"O}zg{\"o}bek, and Su}]{gulla2017adressa}
Jon~Atle Gulla, Lemei Zhang, Peng Liu, {\"O}zlem {\"O}zg{\"o}bek, and Xiaomeng Su. 2017.
\newblock \href {https://doi.org/10.1145/3106426.3109436} {The adressa dataset for news recommendation}.
\newblock In \emph{Proceedings of the International Conference on Web Intelligence}, pages 1042--1048.

\bibitem[{Gunel et~al.(2020)Gunel, Du, Conneau, and Stoyanov}]{gunelsupervised}
Beliz Gunel, Jingfei Du, Alexis Conneau, and Veselin Stoyanov. 2020.
\newblock \href {https://arxiv.org/pdf/2011.01403.pdf} {Supervised contrastive learning for pre-trained language model fine-tuning}.
\newblock In \emph{International Conference on Learning Representations}.

\bibitem[{Heitz et~al.(2022)Heitz, Lischka, Birrer, Paudel, Tolmeijer, Laugwitz, and Bernstein}]{heitz2022benefits}
Lucien Heitz, Juliane~A Lischka, Alena Birrer, Bibek Paudel, Suzanne Tolmeijer, Laura Laugwitz, and Abraham Bernstein. 2022.
\newblock \href {https://doi.org/10.1080/21670811.2021.2021804} {Benefits of diverse news recommendations for democracy: A user study}.
\newblock \emph{Digital Journalism}, 10(10):1710--1730.

\bibitem[{Hu et~al.(2021)Hu, Shen, Wallis, Allen-Zhu, Li, Wang, Wang, and Chen}]{hu2021lora}
Edward~J Hu, Yelong Shen, Phillip Wallis, Zeyuan Allen-Zhu, Yuanzhi Li, Shean Wang, Lu~Wang, and Weizhu Chen. 2021.
\newblock \href {https://arxiv.org/pdf/2106.09685)} {Lora: Low-rank adaptation of large language models}.
\newblock \emph{arXiv preprint arXiv:2106.09685}.

\bibitem[{Hu et~al.(2020)Hu, Xu, Li, Yang, Shi, Duan, Xie, and Zhou}]{hu2020graph}
Linmei Hu, Siyong Xu, Chen Li, Cheng Yang, Chuan Shi, Nan Duan, Xing Xie, and Ming Zhou. 2020.
\newblock \href {https://doi.org/10.18653/v1/2020.acl-main.392} {Graph neural news recommendation with unsupervised preference disentanglement}.
\newblock In \emph{Proceedings of the 58th annual meeting of the association for computational linguistics}, pages 4255--4264.

\bibitem[{Huang et~al.(2013)Huang, He, Gao, Deng, Acero, and Heck}]{huang2013learning}
Po-Sen Huang, Xiaodong He, Jianfeng Gao, Li~Deng, Alex Acero, and Larry Heck. 2013.
\newblock \href {https://doi.org/10.1145/2505515.2505665} {Learning deep structured semantic models for web search using clickthrough data}.
\newblock In \emph{Proceedings of the 22nd ACM international conference on Information \& Knowledge Management}, pages 2333--2338.

\bibitem[{Iana et~al.(2024)Iana, Alam, and Paulheim}]{iana2022survey}
Andreea Iana, Mehwish Alam, and Heiko Paulheim. 2024.
\newblock \href {https://doi.org/10.3233/SW-222991} {A survey on knowledge-aware news recommender systems}.
\newblock \emph{Semantic Web}, 15(1):21--82.

\bibitem[{Iana et~al.(2023{\natexlab{a}})Iana, Glava{\v{s}}, and Paulheim}]{iana2023newsreclib}
Andreea Iana, Goran Glava{\v{s}}, and Heiko Paulheim. 2023{\natexlab{a}}.
\newblock \href {https://doi.org/10.18653/v1/2023.emnlp-demo.26} {Newsreclib: A pytorch-lightning library for neural news recommendation}.
\newblock In \emph{Proceedings of the 2023 Conference on Empirical Methods in Natural Language Processing: System Demonstrations}, pages 296--310.

\bibitem[{Iana et~al.(2023{\natexlab{b}})Iana, Glavas, and Paulheim}]{iana2023simplifying}
Andreea Iana, Goran Glavas, and Heiko Paulheim. 2023{\natexlab{b}}.
\newblock \href {https://doi.org/10.1145/3539618.3592062} {Simplifying content-based neural news recommendation: On user modeling and training objectives}.
\newblock In \emph{Proceedings of the 46th International ACM SIGIR Conference on Research and Development in Information Retrieval}, pages 2384--2388.

\bibitem[{Khosla et~al.(2020)Khosla, Teterwak, Wang, Sarna, Tian, Isola, Maschinot, Liu, and Krishnan}]{khosla2020supervised}
Prannay Khosla, Piotr Teterwak, Chen Wang, Aaron Sarna, Yonglong Tian, Phillip Isola, Aaron Maschinot, Ce~Liu, and Dilip Krishnan. 2020.
\newblock \href {https://dl.acm.org/doi/abs/10.5555/3495724.3497291} {Supervised contrastive learning}.
\newblock In \emph{Proceedings of the 34th International Conference on Neural Information Processing Systems}, pages 18661--18673.

\bibitem[{Kingma and Ba(2014)}]{kingma2014adam}
Diederik~P Kingma and Jimmy Ba. 2014.
\newblock \href {https://doi.org/10.48550/arXiv.1412.6980} {Adam: A method for stochastic optimization}.
\newblock \emph{ICLR}.

\bibitem[{Kummervold et~al.(2021)Kummervold, De~la Rosa, Wetjen, and Brygfjeld}]{kummervold2021operationalizing}
Per~E Kummervold, Javier De~la Rosa, Freddy Wetjen, and Svein~Arne Brygfjeld. 2021.
\newblock \href {https://aclanthology.org/2021.nodalida-main.3} {Operationalizing a national digital library: The case for a norwegian transformer model}.
\newblock In \emph{Proceedings of the 23rd Nordic Conference on Computational Linguistics (NoDaLiDa)}, pages 20--29.

\bibitem[{Li et~al.(2022)Li, Zhu, Bi, Cai, Shang, Dong, Jiang, and Liu}]{li2022miner}
Jian Li, Jieming Zhu, Qiwei Bi, Guohao Cai, Lifeng Shang, Zhenhua Dong, Xin Jiang, and Qun Liu. 2022.
\newblock \href {https://doi.org/10.18653/v1/2022.findings-acl.29} {Miner: Multi-interest matching network for news recommendation}.
\newblock In \emph{Findings of the Association for Computational Linguistics: ACL 2022}, pages 343--352.

\bibitem[{Li and Wang(2019)}]{li2019survey}
Miaomiao Li and Licheng Wang. 2019.
\newblock \href {https://doi.org/10.1109/ACCESS.2019.2944927} {A survey on personalized news recommendation technology}.
\newblock \emph{IEEE Access}, 7:145861--145879.

\bibitem[{Liu et~al.(2020)Liu, Lian, Wang, Qiao, Chen, Sun, and Xie}]{liu2020kred}
Danyang Liu, Jianxun Lian, Shiyin Wang, Ying Qiao, Jiun-Hung Chen, Guangzhong Sun, and Xing Xie. 2020.
\newblock \href {https://doi.org/10.1145/3383313.3412237} {{KRED}: Knowledge-aware document representation for news recommendations}.
\newblock In \emph{Proceedings of the 14th ACM Conference on Recommender Systems}, pages 200--209.

\bibitem[{Liu et~al.(2021)Liu, Shivaram, Culotta, Shapiro, and Bilgic}]{liu2021interaction}
Ping Liu, Karthik Shivaram, Aron Culotta, Matthew~A Shapiro, and Mustafa Bilgic. 2021.
\newblock \href {https://doi.org/10.1145/3442381.3450113} {The interaction between political typology and filter bubbles in news recommendation algorithms}.
\newblock In \emph{Proceedings of the Web Conference 2021}, pages 3791--3801.

\bibitem[{Liu et~al.(2019)Liu, Ott, Goyal, Du, Joshi, Chen, Levy, Lewis, Zettlemoyer, and Stoyanov}]{liu2019roberta}
Yinhan Liu, Myle Ott, Naman Goyal, Jingfei Du, Mandar Joshi, Danqi Chen, Omer Levy, Mike Lewis, Luke Zettlemoyer, and Veselin Stoyanov. 2019.
\newblock \href {https://doi.org/10.48550/arXiv.1907.11692} {{RoBERTa}: A robustly optimized bert pretraining approach}.
\newblock \emph{arXiv preprint arXiv:1907.11692}.

\bibitem[{Lu et~al.(2020)Lu, Dumitrache, and Graus}]{lu2020beyond}
Feng Lu, Anca Dumitrache, and David Graus. 2020.
\newblock \href {https://doi.org/10.1145/3340631.3394864} {Beyond optimizing for clicks: Incorporating editorial values in news recommendation}.
\newblock In \emph{Proceedings of the 28th ACM conference on user modeling, adaptation and personalization}, pages 145--153.

\bibitem[{Ludmann(2017)}]{ludmann2017recommending}
Cornelius~A Ludmann. 2017.
\newblock \href {https://ceur-ws.org/Vol-1866/paper_111.pdf} {Recommending news articles in the clef news recommendation evaluation lab with the data stream management system odysseus.}
\newblock In \emph{CLEF (Working Notes)}.

\bibitem[{Nielsen(2023)}]{nielsen2023scandeval}
Dan~Saattrup Nielsen. 2023.
\newblock \href {https://aclanthology.org/2023.nodalida-1.20} {{ScandEval}: A benchmark for scandinavian natural language processing}.
\newblock In \emph{The 24rd Nordic Conference on Computational Linguistics}.

\bibitem[{Okura et~al.(2017)Okura, Tagami, Ono, and Tajima}]{okura2017embedding}
Shumpei Okura, Yukihiro Tagami, Shingo Ono, and Akira Tajima. 2017.
\newblock \href {https://doi.org/10.1145/3097983.3098108} {Embedding-based news recommendation for millions of users}.
\newblock In \emph{Proceedings of the 23rd ACM SIGKDD international conference on knowledge discovery and data mining}, pages 1933--1942.

\bibitem[{Pariser(2011)}]{pariser2011filter}
Eli Pariser. 2011.
\newblock \emph{The filter bubble: What the Internet is hiding from you}.
\newblock penguin UK.

\bibitem[{Qi et~al.(2021{\natexlab{a}})Qi, Wu, Wu, and Huang}]{qi2021personalized}
Tao Qi, Fangzhao Wu, Chuhan Wu, and Yongfeng Huang. 2021{\natexlab{a}}.
\newblock \href {https://doi.org/10.1145/3404835.3462861} {Personalized news recommendation with knowledge-aware interactive matching}.
\newblock In \emph{Proceedings of the 44th International ACM SIGIR Conference on Research and Development in Information Retrieval}, pages 61--70.

\bibitem[{Qi et~al.(2021{\natexlab{b}})Qi, Wu, Wu, and Huang}]{qi2021pp}
Tao Qi, Fangzhao Wu, Chuhan Wu, and Yongfeng Huang. 2021{\natexlab{b}}.
\newblock \href {https://doi.org/10.18653/v1/2021.acl-long.424} {{PP-Rec}: News recommendation with personalized user interest and time-aware news popularity}.
\newblock In \emph{Proceedings of the 59th Annual Meeting of the Association for Computational Linguistics and the 11th International Joint Conference on Natural Language Processing (Volume 1: Long Papers)}, pages 5457--5467.

\bibitem[{Qi et~al.(2022)Qi, Wu, Wu, and Huang}]{qi2022news}
Tao Qi, Fangzhao Wu, Chuhan Wu, and Yongfeng Huang. 2022.
\newblock \href {https://doi.org/10.1145/3477495.3531778} {News recommendation with candidate-aware user modeling}.
\newblock In \emph{Proceedings of the 45th International ACM SIGIR Conference on Research and Development in Information Retrieval}, pages 1917--1921.

\bibitem[{Qi et~al.(2021{\natexlab{c}})Qi, Wu, Wu, Yang, Yu, Xie, and Huang}]{qi2021hierec}
Tao Qi, Fangzhao Wu, Chuhan Wu, Peiru Yang, Yang Yu, Xing Xie, and Yongfeng Huang. 2021{\natexlab{c}}.
\newblock \href {https://doi.org/10.18653/v1/2021.acl-long.423} {{HieRec}: Hierarchical user interest modeling for personalized news recommendation}.
\newblock In \emph{Proceedings of the 59th Annual Meeting of the Association for Computational Linguistics and the 11th International Joint Conference on Natural Language Processing (Volume 1: Long Papers)}, pages 5446--5456.

\bibitem[{Rao et~al.(2013)Rao, Jia, Feng, and Zhao}]{rao2013taxonomy}
Junyang Rao, Aixia Jia, Yansong Feng, and Dongyan Zhao. 2013.
\newblock \href {https://doi.org/10.1007/978-3-642-41230-1_18} {Taxonomy based personalized news recommendation: novelty and diversity}.
\newblock In \emph{Web Information Systems Engineering--WISE 2013: 14th International Conference, Nanjing, China, October 13-15, 2013, Proceedings, Part I 14}, pages 209--218. Springer.

\bibitem[{Raza(2023)}]{raza2023bias}
Shaina Raza. 2023.
\newblock \href {https://doi.org/10.3390/digital4010003} {Bias reduction news recommendation system}.
\newblock \emph{Digital}, 4(1):92--103.

\bibitem[{Reimers and Gurevych(2019)}]{reimers2019sentence}
Nils Reimers and Iryna Gurevych. 2019.
\newblock \href {https://doi.org/10.18653/v1/D19-1410} {{Sentence-BERT}: Sentence embeddings using siamese bert-networks}.
\newblock In \emph{Proceedings of the 2019 Conference on Empirical Methods in Natural Language Processing and the 9th International Joint Conference on Natural Language Processing (EMNLP-IJCNLP)}, pages 3982--3992.

\bibitem[{Sanh et~al.(2019)Sanh, Debut, Chaumond, and Wolf}]{sanh2019distilbert}
Victor Sanh, Lysandre Debut, Julien Chaumond, and Thomas Wolf. 2019.
\newblock \href {https://arxiv.org/pdf/1910.01108.pdf} {Distilbert, a distilled version of bert: smaller, faster, cheaper and lighter}.
\newblock \emph{arXiv preprint arXiv:1910.01108}.

\bibitem[{Sertkan and Neidhardt(2022)}]{sertkan2022exploring}
Mete Sertkan and Julia Neidhardt. 2022.
\newblock \href {https://doi.org/10.1145/3511047.3536414} {Exploring expressed emotions for neural news recommendation}.
\newblock In \emph{Adjunct Proceedings of the 30th ACM Conference on User Modeling, Adaptation and Personalization}, pages 22--28.

\bibitem[{Sertkan and Neidhardt(2023)}]{sertkan2023effect}
Mete Sertkan and Julia Neidhardt. 2023.
\newblock \href {https://ceur-ws.org/Vol-3561/paper5.pdf} {On the effect of incorporating expressed emotions in news articles on diversity within recommendation models}.
\newblock \emph{decision-making}, 3:11.

\bibitem[{Sheu and Li(2020)}]{sheu2020context}
Heng-Shiou Sheu and Sheng Li. 2020.
\newblock \href {https://doi.org/10.1145/3383313.3418477} {Context-aware graph embedding for session-based news recommendation}.
\newblock In \emph{Proceedings of the 14th ACM Conference on Recommender Systems}, pages 657--662.

\bibitem[{Shi et~al.(2022)Shi, Wang, and Zhai}]{shi2022dcan}
Hao Shi, Zi-Jiao Wang, and Lan-Ru Zhai. 2022.
\newblock \href {https://doi.org/10.48550/arXiv.2206.02627} {{DCAN}: Diversified news recommendation with coverage-attentive networks}.
\newblock \emph{arXiv preprint arXiv:2206.02627}.

\bibitem[{Son et~al.(2013)Son, Kim, and Park}]{son2013location}
Jeong-Woo Son, A-Yeong Kim, and Seong-Bae Park. 2013.
\newblock \href {https://doi.org/10.1145/2484028.2484064} {A location-based news article recommendation with explicit localized semantic analysis}.
\newblock In \emph{Proceedings of the 36th international ACM SIGIR conference on Research and development in information retrieval}, pages 293--302.

\bibitem[{Van~der Maaten and Hinton(2008)}]{van2008visualizing}
Laurens Van~der Maaten and Geoffrey Hinton. 2008.
\newblock \href {https://www.jmlr.org/papers/volume9/vandermaaten08a/vandermaaten08a.pdf?fbcl} {Visualizing data using t-sne.}
\newblock \emph{Journal of machine learning research}, 9(11).

\bibitem[{Vaswani et~al.(2017)Vaswani, Shazeer, Parmar, Uszkoreit, Jones, Gomez, Kaiser, and Polosukhin}]{vaswani2017attention}
Ashish Vaswani, Noam Shazeer, Niki Parmar, Jakob Uszkoreit, Llion Jones, Aidan~N Gomez, {\L}ukasz Kaiser, and Illia Polosukhin. 2017.
\newblock \href {https://dl.acm.org/doi/abs/10.5555/3295222.3295349} {Attention is all you need}.
\newblock In \emph{Proceedings of the 31st International Conference on Neural Information Processing Systems}, pages 6000--6010.

\bibitem[{Vrijenhoek et~al.(2023)Vrijenhoek, B{\'e}n{\'e}dict, Granada, and Odijk}]{vrijenhoek2023radio}
Sanne Vrijenhoek, Gabriel B{\'e}n{\'e}dict, Mateo~Gutierrez Granada, and Daan Odijk. 2023.
\newblock \href {https://doi.org/10.1145/3636465} {Radio*--an introduction to measuring normative diversity in news recommendations}.
\newblock \emph{ACM Transactions on Recommender Systems}.

\bibitem[{Wang et~al.(2018)Wang, Zhang, Xie, and Guo}]{wang2018dkn}
Hongwei Wang, Fuzheng Zhang, Xing Xie, and Minyi Guo. 2018.
\newblock \href {https://doi.org/10.1145/3178876.3186175} {{DKN}: Deep knowledge-aware network for news recommendation}.
\newblock In \emph{Proceedings of the 2018 world wide web conference}, pages 1835--1844.

\bibitem[{Wang et~al.(2022)Wang, Wang, Lu, and Peng}]{wang2022news}
Rongyao Wang, Shoujin Wang, Wenpeng Lu, and Xueping Peng. 2022.
\newblock \href {https://doi.org/10.1109/ICASSP43922.2022.9747149} {News recommendation via multi-interest news sequence modelling}.
\newblock In \emph{ICASSP 2022-2022 IEEE International Conference on Acoustics, Speech and Signal Processing (ICASSP)}, pages 7942--7946. IEEE.

\bibitem[{Wu et~al.(2019{\natexlab{a}})Wu, Wu, An, Huang, Huang, and Xie}]{wu2019naml}
Chuhan Wu, Fangzhao Wu, Mingxiao An, Jianqiang Huang, Yongfeng Huang, and Xing Xie. 2019{\natexlab{a}}.
\newblock \href {https://doi.org/10.24963/ijcai.2019/536} {Neural news recommendation with attentive multi-view learning}.
\newblock In \emph{Proceedings of the 28th International Joint Conference on Artificial Intelligence}, pages 3863--3869.

\bibitem[{Wu et~al.(2019{\natexlab{b}})Wu, Wu, An, Huang, Huang, and Xie}]{wu2019npa}
Chuhan Wu, Fangzhao Wu, Mingxiao An, Jianqiang Huang, Yongfeng Huang, and Xing Xie. 2019{\natexlab{b}}.
\newblock \href {https://doi.org/10.1145/3292500.3330665} {{NPA}: neural news recommendation with personalized attention}.
\newblock In \emph{Proceedings of the 25th ACM SIGKDD international conference on knowledge discovery \& data mining}, pages 2576--2584.

\bibitem[{Wu et~al.(2019{\natexlab{c}})Wu, Wu, An, Huang, and Xie}]{wu2019tanr}
Chuhan Wu, Fangzhao Wu, Mingxiao An, Yongfeng Huang, and Xing Xie. 2019{\natexlab{c}}.
\newblock \href {https://doi.org/10.18653/v1/P19-1110} {Neural news recommendation with topic-aware news representation}.
\newblock In \emph{Proceedings of the 57th Annual meeting of the association for computational linguistics}, pages 1154--1159.

\bibitem[{Wu et~al.(2019{\natexlab{d}})Wu, Wu, Ge, Qi, Huang, and Xie}]{wu2019nrms}
Chuhan Wu, Fangzhao Wu, Suyu Ge, Tao Qi, Yongfeng Huang, and Xing Xie. 2019{\natexlab{d}}.
\newblock \href {https://doi.org/10.18653/v1/D19-1671} {Neural news recommendation with multi-head self-attention}.
\newblock In \emph{Proceedings of the 2019 conference on empirical methods in natural language processing and the 9th international joint conference on natural language processing (EMNLP-IJCNLP)}, pages 6389--6394.

\bibitem[{Wu et~al.(2022{\natexlab{a}})Wu, Wu, and Huang}]{ijcai2022infonce}
Chuhan Wu, Fangzhao Wu, and Yongfeng Huang. 2022{\natexlab{a}}.
\newblock \href {https://doi.org/10.24963/ijcai.2022/348} {Rethinking {InfoNCE}: How many negative samples do you need?}
\newblock In \emph{Proceedings of the Thirty-First International Joint Conference on Artificial Intelligence, {IJCAI-22}}, pages 2509--2515. International Joint Conferences on Artificial Intelligence Organization.

\bibitem[{Wu et~al.(2023)Wu, Wu, Huang, and Xie}]{wu2023personalized}
Chuhan Wu, Fangzhao Wu, Yongfeng Huang, and Xing Xie. 2023.
\newblock \href {https://doi.org/10.1145/3530257} {Personalized news recommendation: Methods and challenges}.
\newblock \emph{ACM Transactions on Information Systems}, 41(1):1--50.

\bibitem[{Wu et~al.(2020{\natexlab{a}})Wu, Wu, Qi, and Huang}]{wu2020sentirec}
Chuhan Wu, Fangzhao Wu, Tao Qi, and Yongfeng Huang. 2020{\natexlab{a}}.
\newblock \href {https://aclanthology.org/2020.aacl-main.6} {{SentiRec}: Sentiment diversity-aware neural news recommendation}.
\newblock In \emph{Proceedings of the 1st Conference of the Asia-Pacific Chapter of the Association for Computational Linguistics and the 10th International Joint Conference on Natural Language Processing}, pages 44--53.

\bibitem[{Wu et~al.(2021)Wu, Wu, Qi, and Huang}]{wu2021empowering}
Chuhan Wu, Fangzhao Wu, Tao Qi, and Yongfeng Huang. 2021.
\newblock \href {https://doi.org/10.1145/3404835.3463069} {Empowering news recommendation with pre-trained language models}.
\newblock In \emph{Proceedings of the 44th International ACM SIGIR Conference on Research and Development in Information Retrieval}, pages 1652--1656.

\bibitem[{Wu et~al.(2022{\natexlab{b}})Wu, Wu, Qi, and Huang}]{wu2022end}
Chuhan Wu, Fangzhao Wu, Tao Qi, and Yongfeng Huang. 2022{\natexlab{b}}.
\newblock \href {https://doi.org/10.48550/arXiv.2204.00539} {End-to-end learnable diversity-aware news recommendation}.
\newblock \emph{arXiv preprint arXiv:2204.00539}.

\bibitem[{Wu et~al.(2022{\natexlab{c}})Wu, Wu, Qi, Li, and Huang}]{wu2022news}
Chuhan Wu, Fangzhao Wu, Tao Qi, Chenliang Li, and Yongfeng Huang. 2022{\natexlab{c}}.
\newblock \href {https://doi.org/10.1145/3477495.3531862} {Is news recommendation a sequential recommendation task?}
\newblock In \emph{Proceedings of the 45th International ACM SIGIR Conference on Research and Development in Information Retrieval}, pages 2382--2386.

\bibitem[{Wu et~al.(2022{\natexlab{d}})Wu, Wu, Qi, Zhang, Xie, and Huang}]{wu2022removing}
Chuhan Wu, Fangzhao Wu, Tao Qi, Wei-Qiang Zhang, Xing Xie, and Yongfeng Huang. 2022{\natexlab{d}}.
\newblock \href {https://doi.org/10.1057/s41599-022-01473-1} {Removing ai’s sentiment manipulation of personalized news delivery}.
\newblock \emph{Humanities and Social Sciences Communications}, 9(1):1--9.

\bibitem[{Wu et~al.(2020{\natexlab{b}})Wu, Qiao, Chen, Wu, Qi, Lian, Liu, Xie, Gao, Wu, and Zhou}]{wu2020mind}
Fangzhao Wu, Ying Qiao, Jiun-Hung Chen, Chuhan Wu, Tao Qi, Jianxun Lian, Danyang Liu, Xing Xie, Jianfeng Gao, Winnie Wu, and Mind Zhou. 2020{\natexlab{b}}.
\newblock \href {https://doi.org/10.18653/v1/2020.acl-main.331} {{MIND}: A large-scale dataset for news recommendation}.
\newblock In \emph{Proceedings of the 58th Annual Meeting of the Association for Computational Linguistics}, pages 3597--3606.

\bibitem[{Xu et~al.(2023)Xu, Peng, Liu, Sun, and Wang}]{xu2023group}
Hongyan Xu, Qiyao Peng, Hongtao Liu, Yueheng Sun, and Wenjun Wang. 2023.
\newblock \href {https://doi.org/10.1145/3584946} {Group-based personalized news recommendation with long-and short-term fine-grained matching}.
\newblock \emph{ACM Transactions on Information Systems}.

\bibitem[{Xun et~al.(2021)Xun, Zhang, Zhao, Zhu, Zhang, Li, He, He, Chua, and Wu}]{xun2021we}
Jiahao Xun, Shengyu Zhang, Zhou Zhao, Jieming Zhu, Qi~Zhang, Jingjie Li, Xiuqiang He, Xiaofei He, Tat-Seng Chua, and Fei Wu. 2021.
\newblock \href {https://doi.org/10.1145/3474085.3475514} {Why do we click: visual impression-aware news recommendation}.
\newblock In \emph{Proceedings of the 29th ACM International Conference on Multimedia}, pages 3881--3890.

\bibitem[{Yi et~al.(2021)Yi, Wu, Wu, Liu, Sun, and Xie}]{yi2021efficient}
Jingwei Yi, Fangzhao Wu, Chuhan Wu, Ruixuan Liu, Guangzhong Sun, and Xing Xie. 2021.
\newblock \href {https://doi.org/10.18653/v1/2021.emnlp-main.223} {Efficient-fedrec: Efficient federated learning framework for privacy-preserving news recommendation}.
\newblock In \emph{Proceedings of the 2021 Conference on Empirical Methods in Natural Language Processing}, pages 2814--2824.

\bibitem[{Zhang and Hurley(2008)}]{zhang2008avoiding}
Mi~Zhang and Neil Hurley. 2008.
\newblock \href {https://doi.org/10.1145/1454008.1454030} {Avoiding monotony: improving the diversity of recommendation lists}.
\newblock In \emph{Proceedings of the 2008 ACM conference on Recommender systems}, pages 123--130.

\end{thebibliography}

\appendix
\section{Dataset Statistics}
\label{sec:appendix_datasets}
\newcolumntype{g}{>{\columncolor{Gray}}r}
\begin{table}[h]
\resizebox{\columnwidth}{!}{%
  \begin{tabular}{lrg|rg}
    \toprule
      & \multicolumn{2}{c}{\textbf{MIND (large)}} & \multicolumn{2}{c}{\textbf{Adressa (one week)}}\\ 
     \cmidrule(lr){2-3} \cmidrule(lr){4-5}
    
    \multicolumn{1}{l}{Statistic} & \multicolumn{1}{r}{Train} & \multicolumn{1}{r|}{Test} & \multicolumn{1}{r}{Train} & \multicolumn{1}{r}{Test} \\
    \hline
    
    \# News & 101,527 & 72,023 & 11,207 & 11,207 \\
    \# Users & 698,365 & 196,444 & 96,801 & 68,814 \\
    \# Impressions & 2,186,683 & 365,201 & 218,848 & 146,284 \\ 
    \hdashline
    \# Categories & 18 & 17 & 18 & 18 \\
    \hdashline
    Avg. history length & 33.7 & 33.6 & 13.9 & 15.6 \\
    Avg. \# candidates / user & 37.4 & 37.4 & 21.0 & 21.0 \\
  \bottomrule
\end{tabular}%
}
\caption{MIND and Adressa dataset statistics.}
\label{tab:datasets}
\vspace{-0.5em}
\end{table}
\section{Reproducibility Details}
\label{sec:appendx_reproducibility}

\subsection{Model Parameters.}
\label{sec:appendix_model_parameters}
\newcolumntype{g}{>{\columncolor{Gray}}c}

\begin{table}[h]
\centering
    \resizebox{\columnwidth}{!}{%
    \begin{tabular}{lr|rg|rg}
        \toprule
         & \multicolumn{1}{c}{}
         & \multicolumn{2}{c}{\textbf{MIND}} 
         & \multicolumn{2}{c}{\textbf{Adressa}} 
         \\ \cmidrule(lr){3-4}  \cmidrule(lr){5-6}
         
        \textbf{Model} 
        & Non-trainable
        & Trainable & Total 
        & Trainable & Total
        \\ \hline
        
        NRMS-PLM 
        & 56.7 
        & 73 & 129 
        & 126 & 182 
        \\

        MINER 
        & 56.7 
        & 68.2 & 124 
        & 121 & 178 
        \\ 
        \hdashline
        
        NAML-PLM
        & 56.7
        & 70.8 & 127
        & 124  & 180
        \\
        
        LSTUR-PLM 
        & 56.7 
        & 633 & 690 
        & 200 & 257 
        \\
        
        MINS-PLM 
        & 56.7 
        & 73.3 & 130 
        & 126 & 183 
        \\ 
        
        CAUM\textsubscript{no entities}-PLM
        & 56.7 
        & 73.2 & 129 
        & 126 & 183 
        \\ 

        CAUM-PLM
        & 56.7 
        & 74.9 & 131 
        & -- & --
        \\

        TANR-PLM
        & 56.7 
        & 70.6 & 127 
        & 123 & 180 
        \\

        \hdashline
        
        SentiRec-PLM 
        & 56.7 
        & 73 & 129 
        & 126 & 182 
        \\ 

        SentiDebias-PLM 
        & 56.7 
        & 73.3 & 130 
        & 126 & 183 
        \\ 
        \hline 
        
        \manner{} (CR-Module\textsubscript{title} / A-Module\textsubscript{title}) -- monolingual 
        & 56.7 
        & 67.9 & 124
        & 121 & 177
        \\

        \manner{} (CR-Module / A-Module) -- monolingual 
        & 56.7 
        & 70.3 & 126 
        & -- & --
        \\ 

        \hdashline
         \manner{} (CR-Module\textsubscript{title} / A-Module\textsubscript{title}) -- multilingual 
        & 0
        & 134 & 134
        & 134 & 134
        \\ 
        
        \bottomrule
    \end{tabular}%
}
\caption{Number of model parameters (in millions). CR-Module\textsubscript{title} / A-Module\textsubscript{title} denote the \manner{} modules trained with only the news title as input to the NE.}
\label{tab:model_parameters}
\vspace{-0.5em}
\end{table}

\subsection{Hyperparameters and Implementation}
\label{sec:appendix_hyperparameters}

\vspace{1.4mm}
\noindent\textbf{Hyperparameter Optimization.}
We use RoBERTa Base \cite{liu2019roberta} and NB-BERT Base \cite{kummervold2021operationalizing,nielsen2023scandeval} as the backbone PLMs of all models, in experiments on MIND and Adressa, respectively. In both cases, we fine-tune only the PLM's last four layers.\footnote{In preliminary results, we did not see significant differences between full fine-tuning of all layers and fine-tuning only the last four layers. In the interest of computational efficiency, we thus froze the first eight layers of the transformer.}
In the cross-lingual transfer experiments from \$\ref{sec:xlt}, we fine-tune all of the 6 layers of DistilBERT.
We use 100-dimensional TransE embeddings \cite{bordes2013translating} pretrained on Wikidata as input to the entity encoder in the NE of the knowledge-aware NNRs. We perform hyperparameter tuning on the main hyperparameters of \manner{} and the baselines using grid search. Table \ref{tab:hyperparameters} lists the search spaces for all the optimized hyperparameters, as well as the best values. We repeat each experiment five times with the seeds ($\{42, 43, 44, 45, 46\}$) set with PyTorch Lightning's \texttt{seed\_everything}.
\begin{table*}[ht]
\centering

\resizebox{\textwidth}{!}{%
    \begin{tabular}{lccccccccccc} 
    \toprule
     &  \multicolumn{1}{c}{\texttt{\texttt{lr}}} &  \multicolumn{1}{c}{\texttt{num\textsubscript{heads}}} &  \multicolumn{1}{c}{\texttt{query\textsubscript{dim}}} &  \multicolumn{1}{c}{\texttt{UE agg}} &  \multicolumn{1}{c}{$K$} &  \multicolumn{1}{c}{\texttt{score agg}} &  \multicolumn{1}{c}{$\lambda$} &  \multicolumn{1}{c}{$\mu$} &  \multicolumn{1}{c}{$\alpha$} &  \multicolumn{1}{c}{$\tau_{CR-Module}$} &  \multicolumn{1}{c}{$\tau_{A-Module}$} \\
        \bottomrule
        \textbf{Search Space} & [$1e^{-4}$, $1e^{-6}$] & \{8, 12, 16, 24, 32\} & [50, 200]  & \{ini, con\} & \{8, 16, 32, 48\} & \{mean, max, weighted\} & [0.1, 0.3] & [5, 15] & [0.05, 0.2] & [0.1, 0.5] & [0.1, 0.9] \\

        \textbf{Step} & $1e^{-1}$ & -- & 50 & -- & -- & -- & 0.05 & 5 & 0.05 & 0.02 & 0.05 \\

        \hline  
        NRMS-PLM & $1e^{-5}$ / $1e^{-6}$ & 32 / 8 &  150 / 200 & -- & -- & -- & -- & -- & -- & -- & -- \\
        MINER & $1e^{-5}$ / $1e^{-6}$ & -- & -- & -- & 32 / 48 & mean / mean & -- & -- & -- & -- & -- \\

        \hdashline 
        
        NAML-PLM & $1e^{-5}$ / $1e^{-6}$ & 16 / 8 & 200 / 200  & -- & -- & -- & -- & -- & -- & -- & -- \\
        LSTUR-PLM & $1e^{-5}$ / $1e^{-6}$ & 24 / 8 & 150 / 50 & ini / ini & -- & -- & -- & -- & -- & -- & -- \\
        MINS-PLM & $1e^{-5}$ / $1e^{-6}$ & 32 / 12 & 100 / 200 & -- & -- & -- & -- & -- & -- & -- & -- \\
        CAUM-PLM & $1e^{-5}$ / $1e^{-6}$ & 16 / 16 & 50 / 150 & -- & -- & -- & -- & -- & -- & -- & -- \\
        TANR-PLM & $1e^{-5}$ / $1e^{-6}$ & 32 / 8 & 150 / 50 & -- & -- & -- & 0.3 / 0.15 & -- & -- & -- & -- \\

        \hdashline
        SentiRec-PLM & $1e^{-5}$ / $1e^{-6}$ & 32 / 8 & 200 / 200 & -- & -- & -- & -- & 5 / 5 & -- & -- & -- \\
        SentiDebias-PLM & $1e^{-5}$ / $1e^{-6}$ & 8 / 12 & 100 / 150 & -- & -- & -- & -- & -- & 0.15 / 0.15 & -- & -- \\

        \hdashline
        
        MANNeR& $1e^{-5}$ / $1e^{-6}$ & -- & 200 / 200 & -- & -- & -- & -- & -- & -- & 0.36 / 0.14 & 0.9 / 0.9  \\
        \bottomrule
    \end{tabular}%
    }

\caption{Search spaces used for hyperparameter optimization and best values found for all models. We report the optimal values in the format \textit{value\textsubscript{MIND} / value\textsubscript{Adressa}}. We use the following abbreviations: \texttt{lr} = learning rate, \texttt{num\textsubscript{heads}} = number of attention heads, \texttt{query\textsubscript{dim}} = dimensionality of the query vector in additive attention, \texttt{UE agg} = aggregation method used to combine the long-term and the short-term user representations into a final user embedding in LSTUR \cite{an2019lstur}, $K$ = number of context codes in MINER \cite{li2022miner}, \texttt{score agg} = aggregation function for the final user click score calculation in MINER \cite{li2022miner}, $\lambda$ = weight of the topic classification task in TANR \cite{wu2019tanr}, $\mu$ = weight of the sentiment diversity regularization loss in SentiRec \cite{wu2020sentirec}, $\alpha$ = adversarial loss coefficient in SentiDebias \cite{wu2022removing}, $\tau$ = temperature parameter in SCL in \manner{}, \textit{ini} = initialize, \textit{con} = concatenate, \textit{categ} = category.}
\label{tab:hyperparameters}
\vspace{-0.5em}
\end{table*}

\vspace{1.4mm}
\noindent\textbf{Code.} We train \manner{}, as well as all the baselines, using the implementations provided in the NewsRecLib library \cite{iana2023newsreclib}.

\vspace{1.4mm}
\noindent\textbf{Infrastructure and Compute.}
We conduct all experiments on a cluster with virtual machines. We train \manner{} on both datasets, as well as the baselines on MIND, on a single NVIDIA A100 40GB GPU. We train the baselines on Adressa on a single NVIDIA Tesla V100 32GB GPU.

\section{Additional Results}
\label{sec:appendix_additional_results}

\subsection{Content Personalization}
\label{sec:appendix_content_personalization}

\begin{figure*}[h]
     \centering
     \begin{subfigure}[b]{\columnwidth}
         \centering
         \includegraphics[width=\columnwidth]{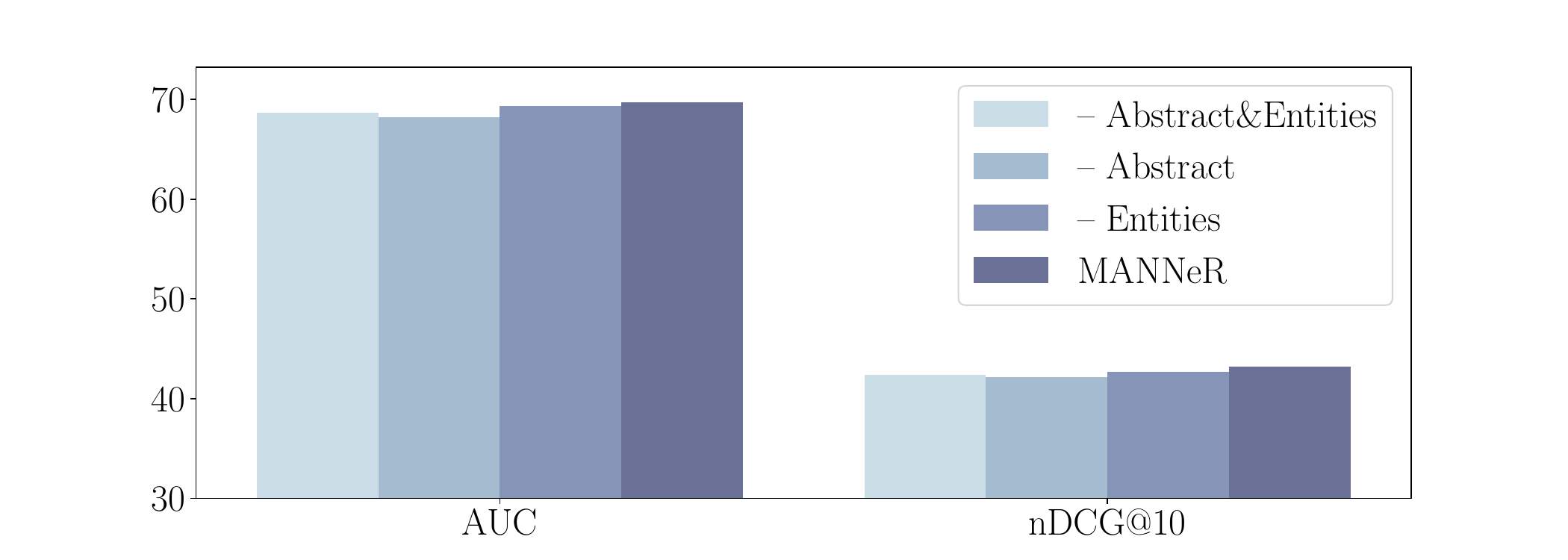}
         \caption{Input features for the News Encoder\,(NE).}
         \label{fig:ablation_features}
     \end{subfigure}
     \hfill
     \begin{subfigure}[b]{\columnwidth}
         \centering
         \includegraphics[width=\columnwidth]{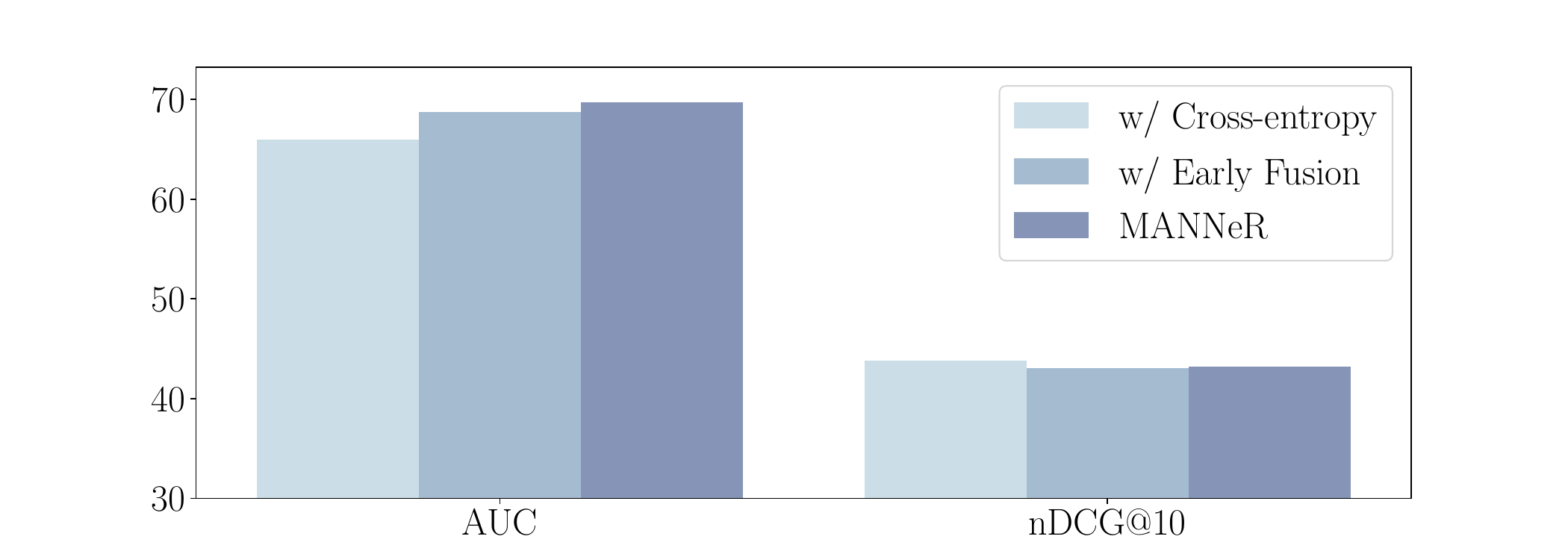}
         \caption{\texttt{CR-Module} design/training alternatives.}
         \label{fig:ablation_training}
     \end{subfigure}
    \caption{Effect of different (a) NE inputs and (b) model design/training choices on \manner{}'s content-based personalization performance.}
    \label{fig:ablation}
    \vspace{-0.5em}
\end{figure*}

Fig.~\ref{fig:ablation_features} shows \manner{}'s performance on MIND for different inputs to the NE. Even the \texttt{CR-Module} exposed to titles only (i.e., no abstract or entity information) outperforms all of the baselines on content recommendation.
Fig. \ref{fig:ablation_training} illustrates \manner{}'s performance for alternative architecture designs and training objectives (cf. \$\ref{sec:content_personalization}).\footnote{For brevity, we report results on MIND; findings on Adressa exhibit identical trends.}

\subsection{Single-Aspect Customization}
\label{sec:appendix_single_apsect_customization}

\begin{figure*}[h!]
     \centering
     \begin{subfigure}[b]{0.48\textwidth}
         \centering
         \includegraphics[width=\textwidth]{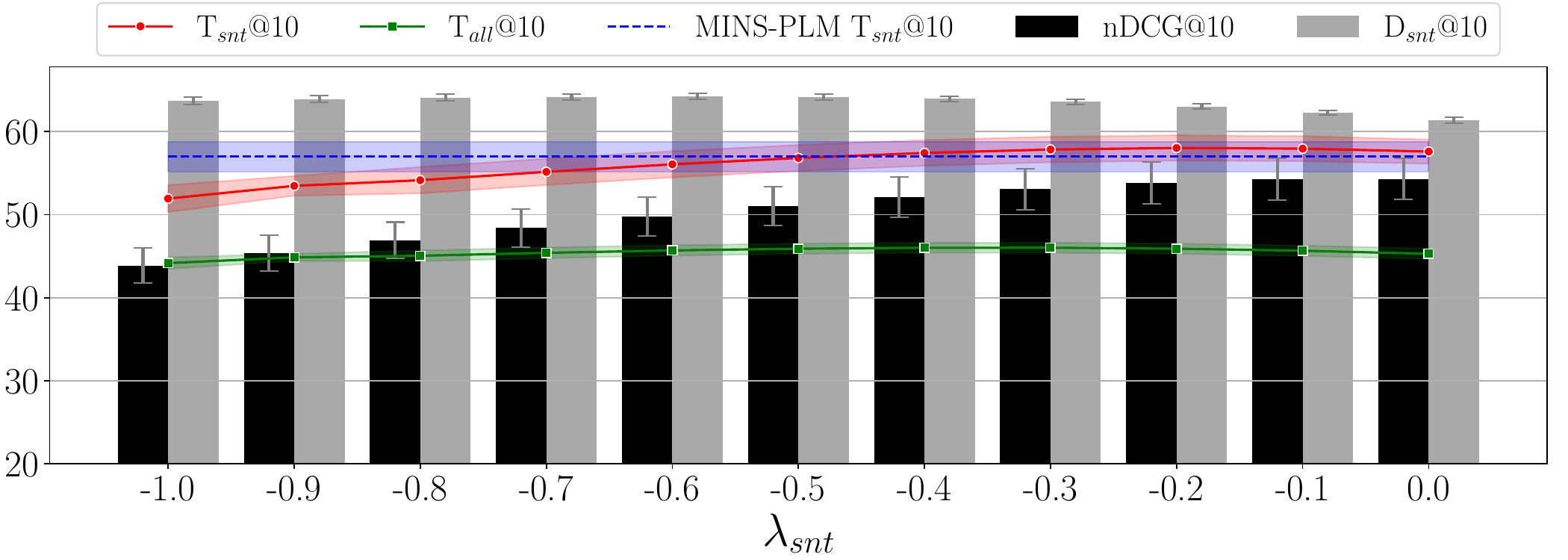}
         \caption{Single-aspect sentiment diversification.}
         \label{fig:single_aspect_div_sent_adressa}
     \end{subfigure}
     \hfill
     \begin{subfigure}[b]{0.48\textwidth}
         \centering
         \includegraphics[width=\textwidth]{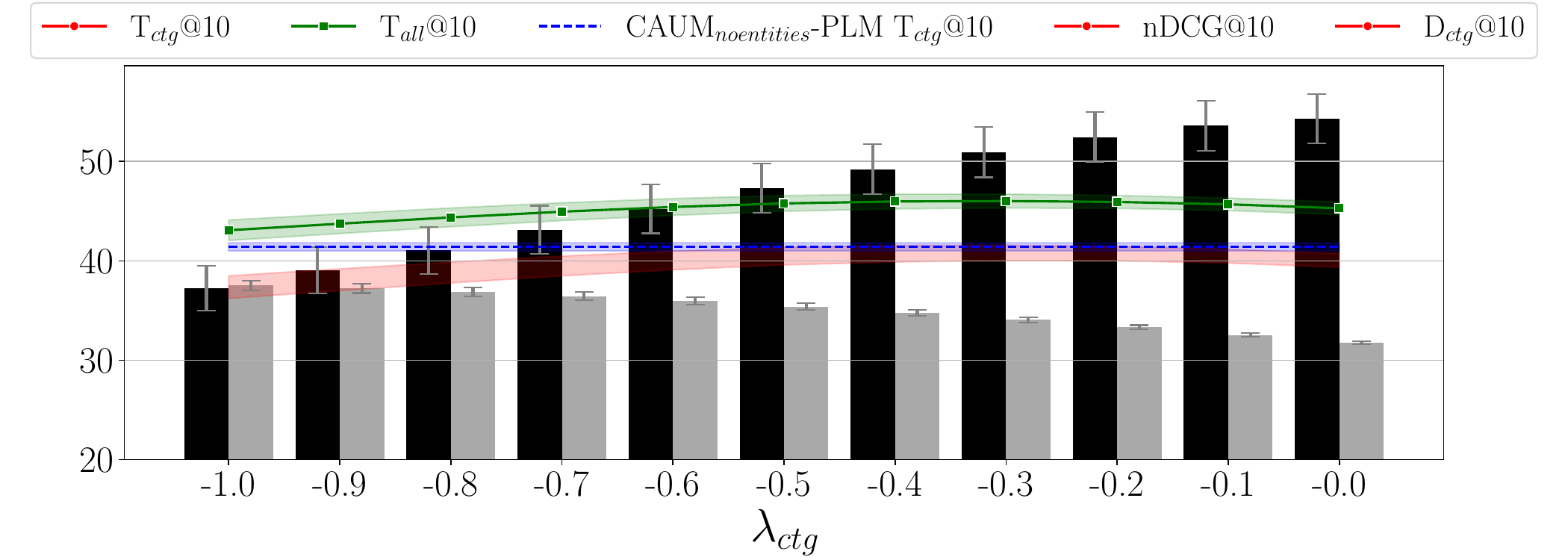}
         \caption{Single-aspect category diversification.}
         \label{fig:single_aspect_div_categ_adressa}
     \end{subfigure}
     \hfill
     \begin{subfigure}[b]{0.48\textwidth}
         \centering
         \includegraphics[width=\textwidth]{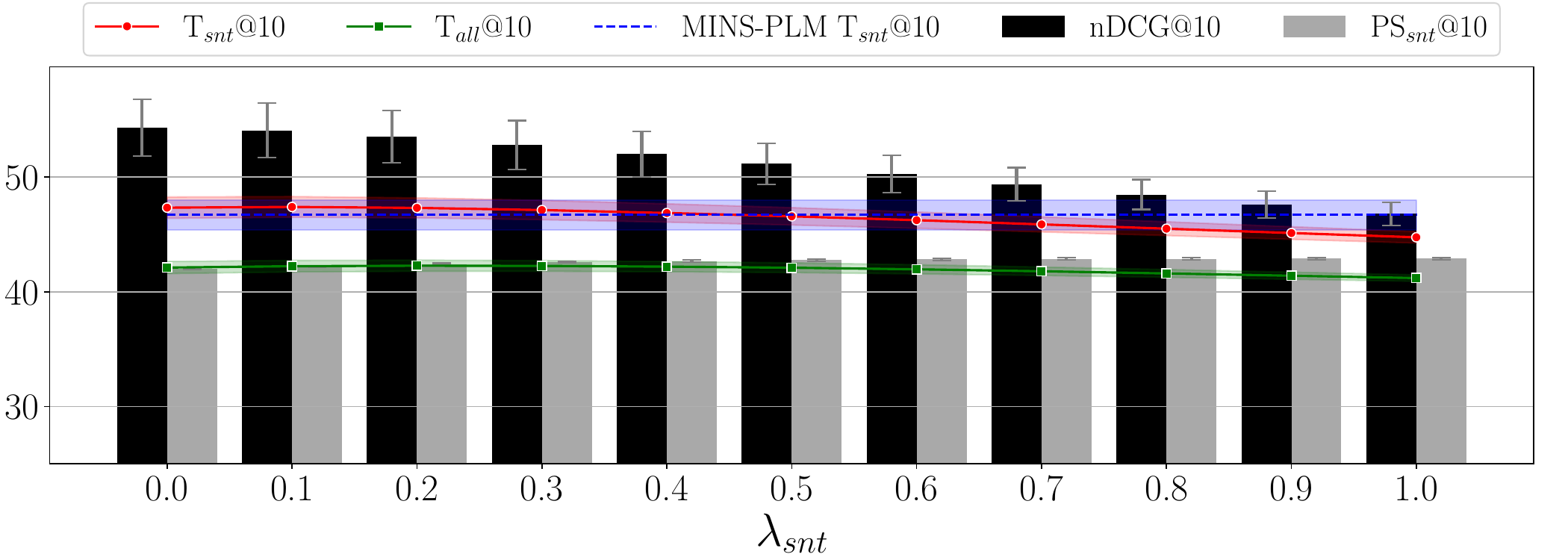}
         \caption{Single-aspect sentiment personalization.}
         \label{fig:single_aspect_pers_sent_adressa}
     \end{subfigure}
     \hfill
     \begin{subfigure}[b]{0.48\textwidth}
         \centering
         \includegraphics[width=\textwidth]{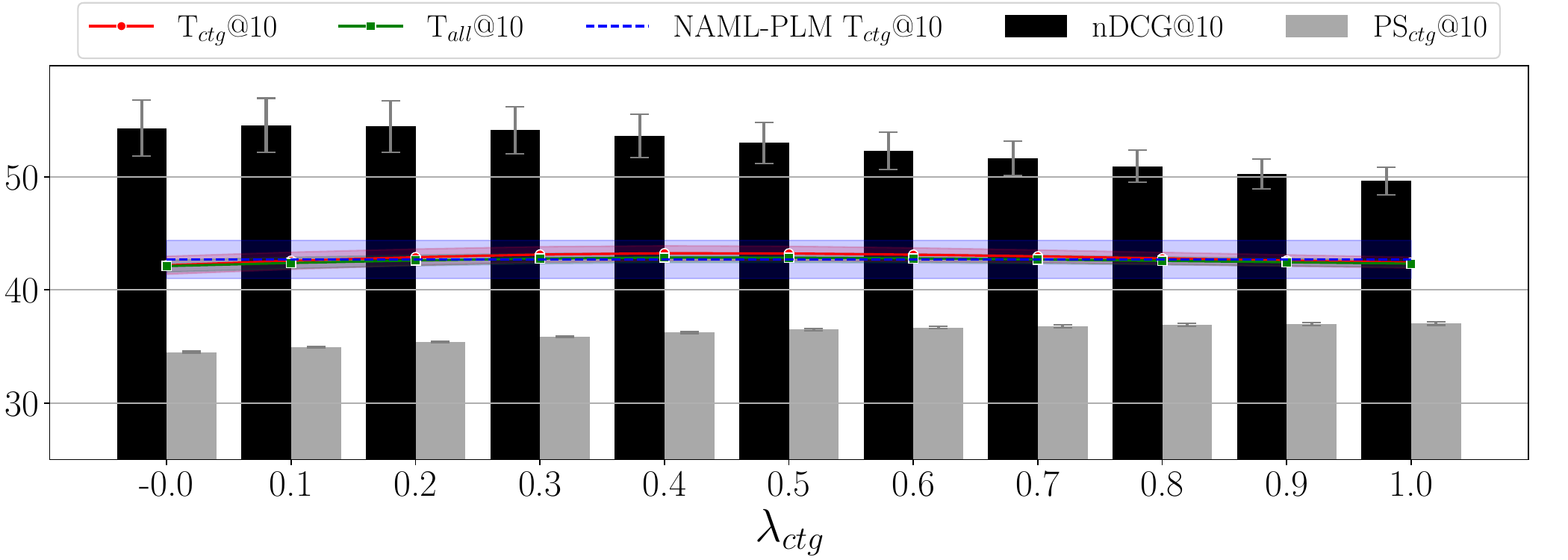}
         \caption{Single-aspect category personalization.}
         \label{fig:single_aspect_pers_categ_adressa}
     \end{subfigure}
    \caption{Results of single-aspect customization for \manner{} and the best baseline on Adressa. 
    }
    \label{fig:single_aspect_results_addresa}
    \vspace{-1em}
\end{figure*}
\begin{figure*}[t]
     \centering
    \begin{subfigure}[b]{0.48\textwidth}
         \centering
         \includegraphics[width=\textwidth]{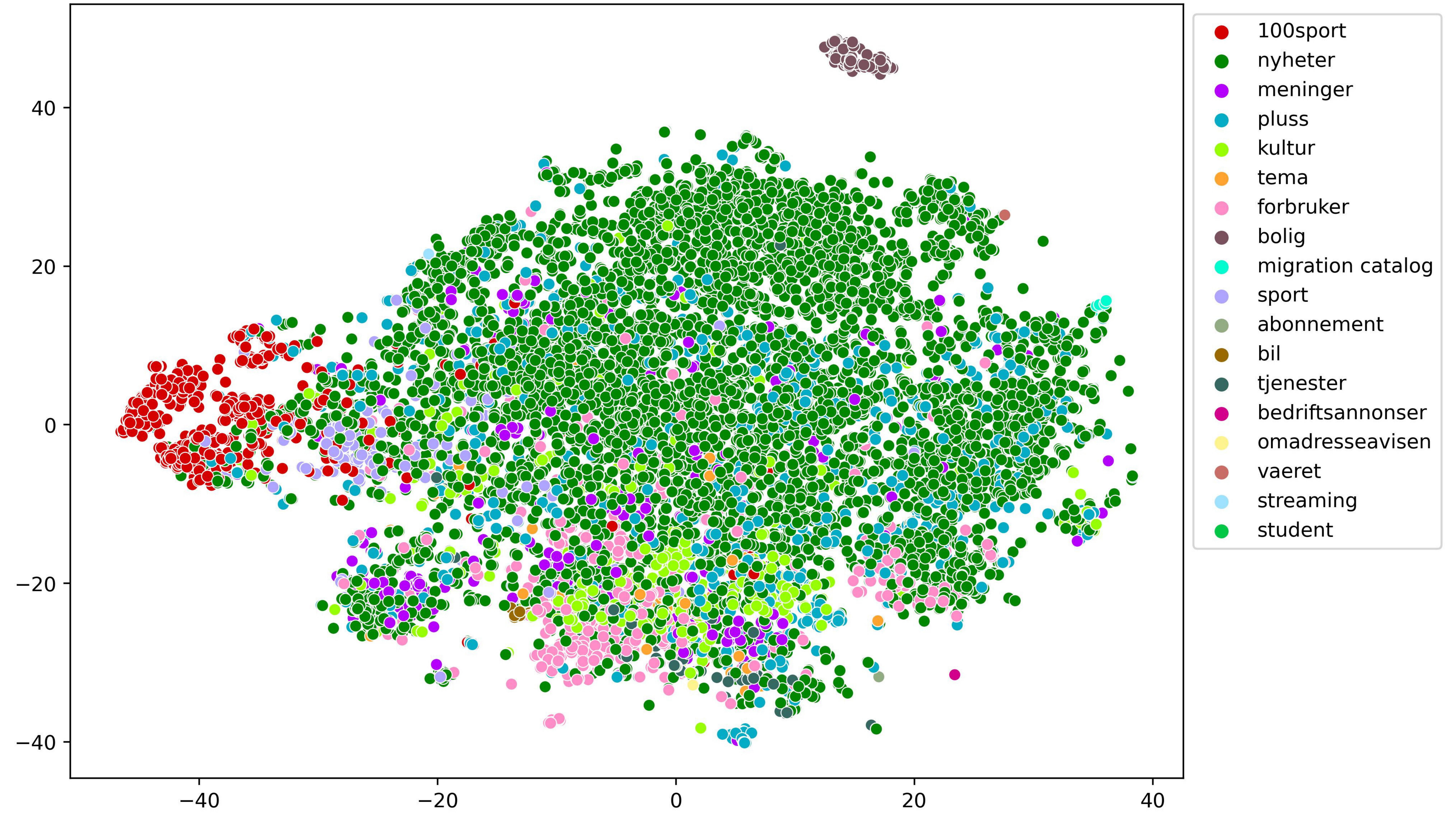}
         \caption{Category-shaped embedding space.}
         \label{fig:tsne_categ_adresa}
     \end{subfigure}
     \hfill
     \begin{subfigure}[b]{0.48\textwidth}
         \centering
         \includegraphics[width=0.95\textwidth]{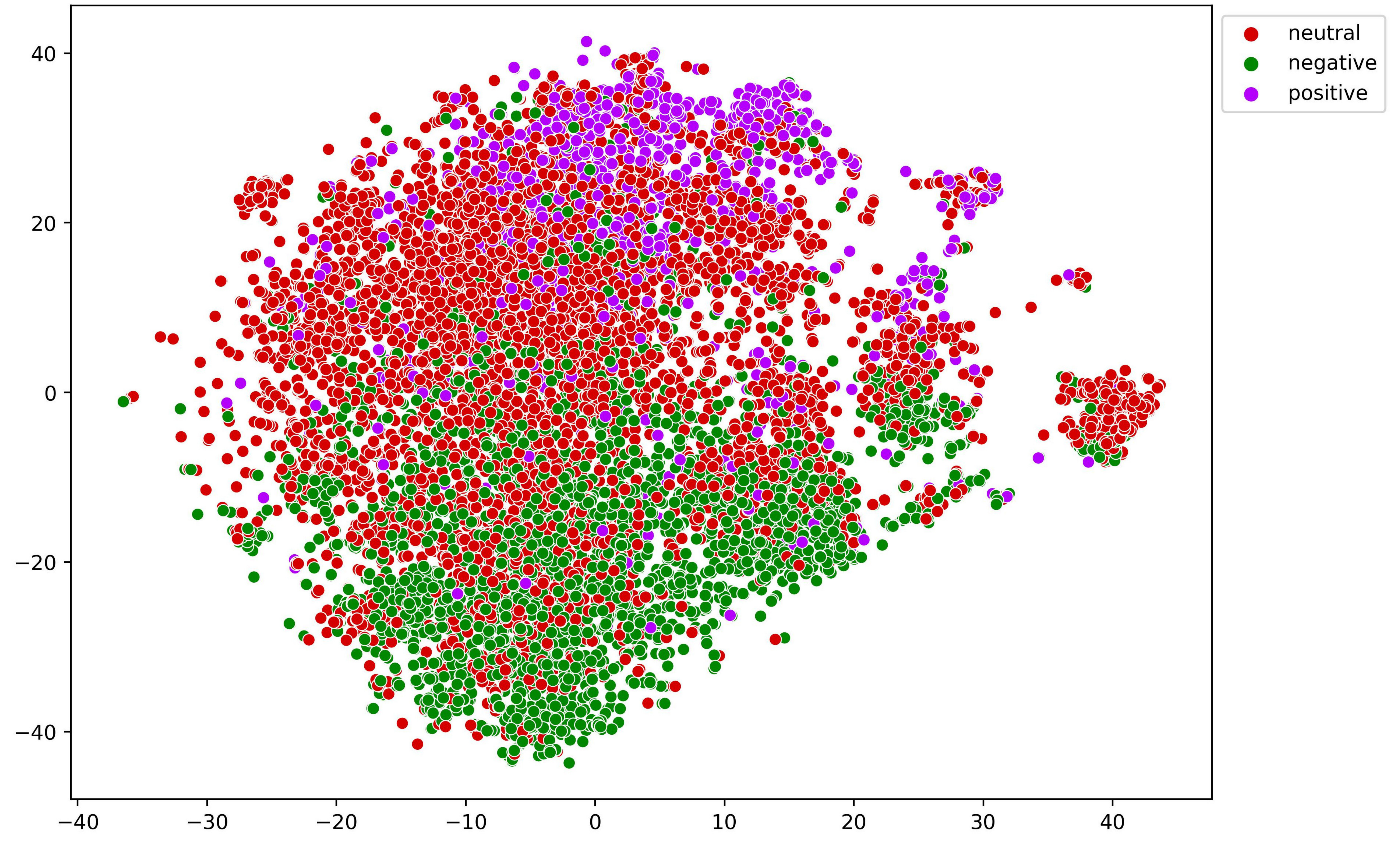}
         \caption{Sentiment-shaped embedding space.}
         \label{fig:tsne_sent_adressa}
     \end{subfigure}
    \caption{t-SNE plots of the news embeddings in the test set of Adressa.}
    \label{fig:tsne_embeddings_adressa}
    \vspace{-1em}
\end{figure*}

Figure \ref{fig:single_aspect_results_addresa} explores the trade-off between content and aspect diversification, and respectively, personalization tasks for different values of $\lambda$\textsubscript{ctg} and $\lambda$\textsubscript{snt} on the Adressa dataset. 
Fig. \ref{fig:tsne_embeddings_adressa} shows the 2-dimensional t-SNE visualizations \cite{van2008visualizing} of the news embeddings produced with aspect-specialized NEs trained on Adressa. 

\subsection{Multi-Aspect Customization}
\label{sec:appendix_multi_apsect_customization}

\begin{figure*}[h!]
     \centering
     \begin{subfigure}[b]{0.48\textwidth}
         \centering
         \includegraphics[width=\textwidth]{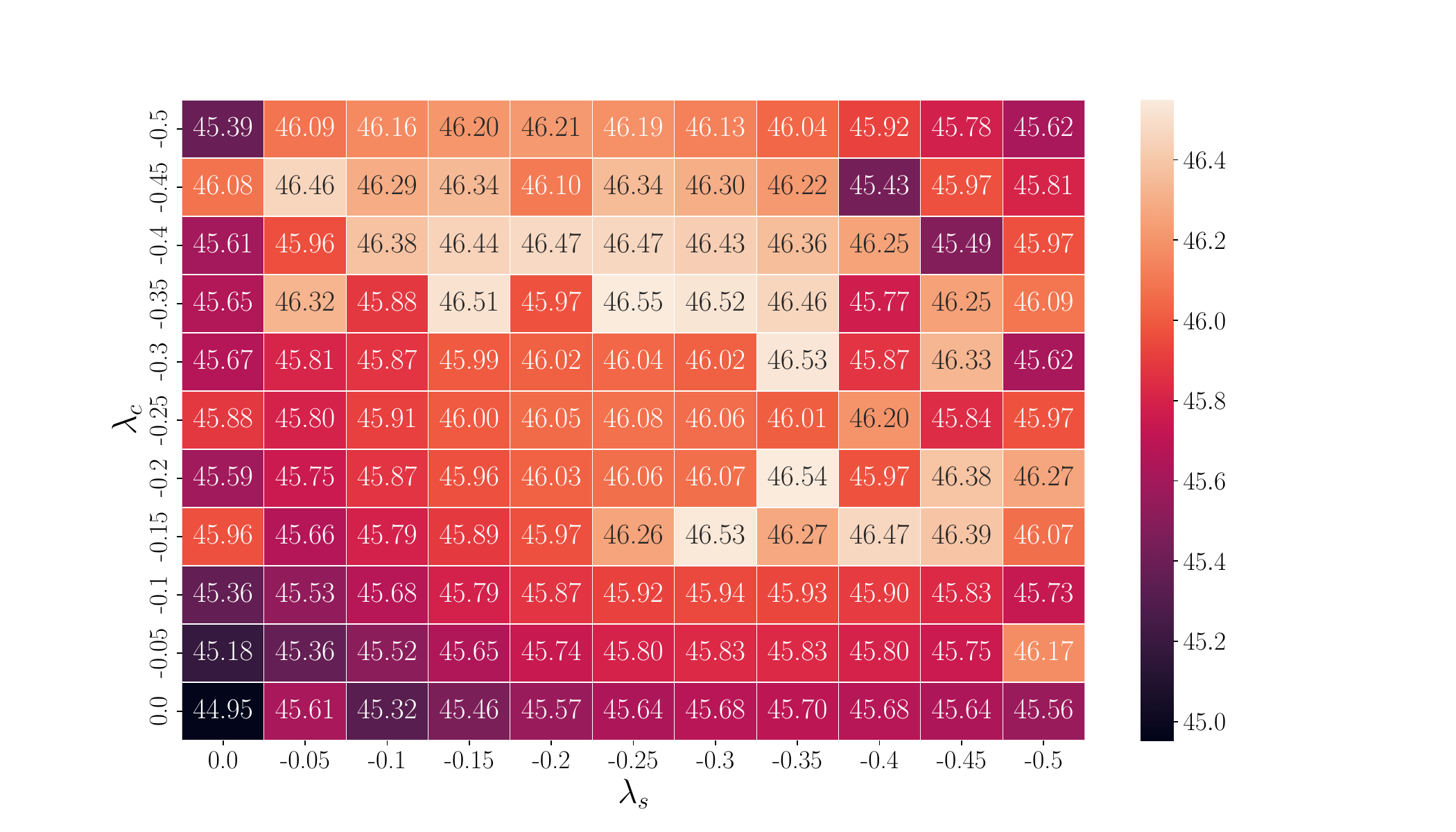}
         \caption{Multi-aspect diversification on Adressa.}
         \label{fig:multi_aspect_div_adressa}
     \end{subfigure}
     \hfill
     \begin{subfigure}[b]{0.48\textwidth}
         \centering
         \includegraphics[width=\textwidth]{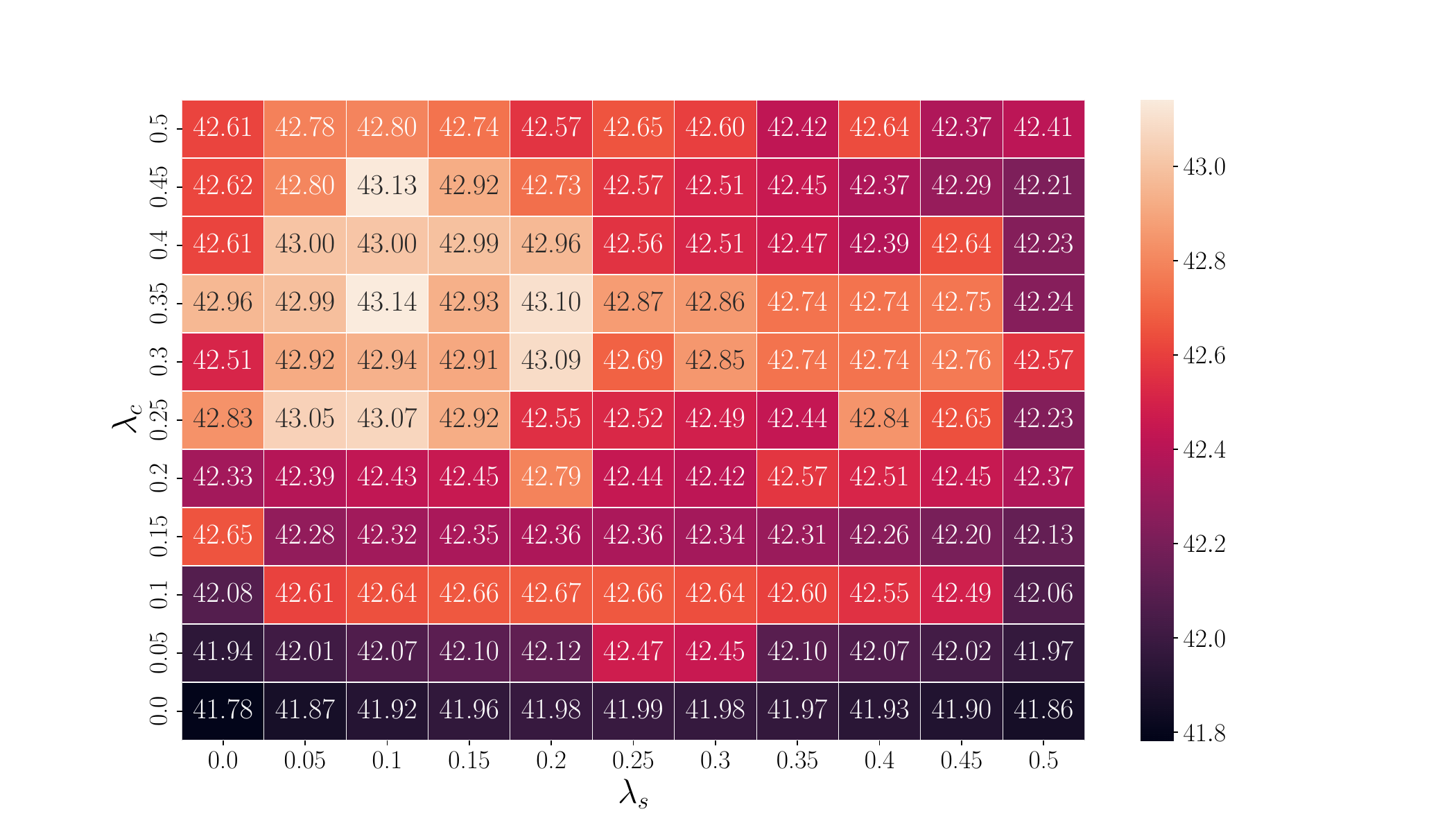}
         \caption{Multi-aspect personalization on Adressa.}
         \label{fig:multi_aspect_pers_adressa}
     \end{subfigure}
    \caption{Results of multi-aspect customization for \manner{} on Adressa.}
    \label{fig:multi_aspect_results_addressa}
    \vspace{-1em}
\end{figure*}

Fig. \ref{fig:multi_aspect_results_addressa} explores the trade-off between content personalization and multi-aspect diversification on Adressa.

\subsection{Cross-Lingual Transfer}
\label{sec:appendix_xlt}

\begin{figure*}[h!]
     \centering
     \begin{subfigure}[b]{0.48\textwidth}
         \centering
         \includegraphics[width=\textwidth]{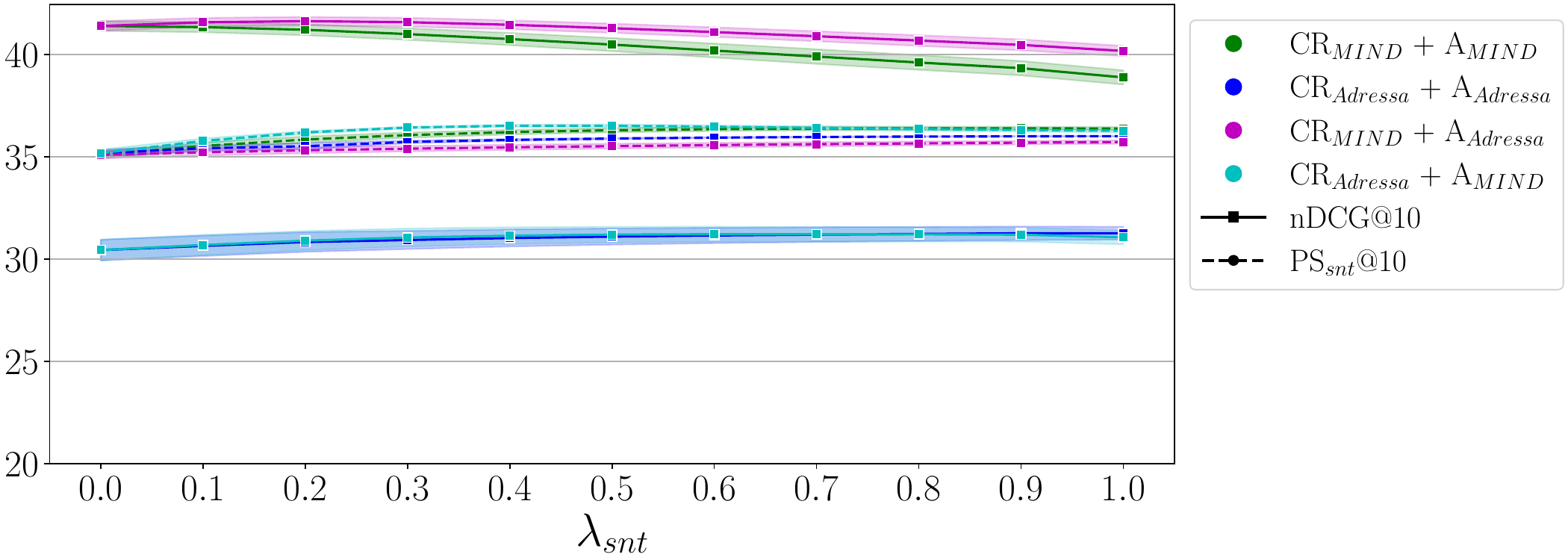}
         \caption{Sentiment personalization.}
         \label{fig:adressa_transfer_mind_pers_sent}
     \end{subfigure}
     \hfill
     \begin{subfigure}[b]{0.48\textwidth}
         \centering
         \includegraphics[width=\textwidth]{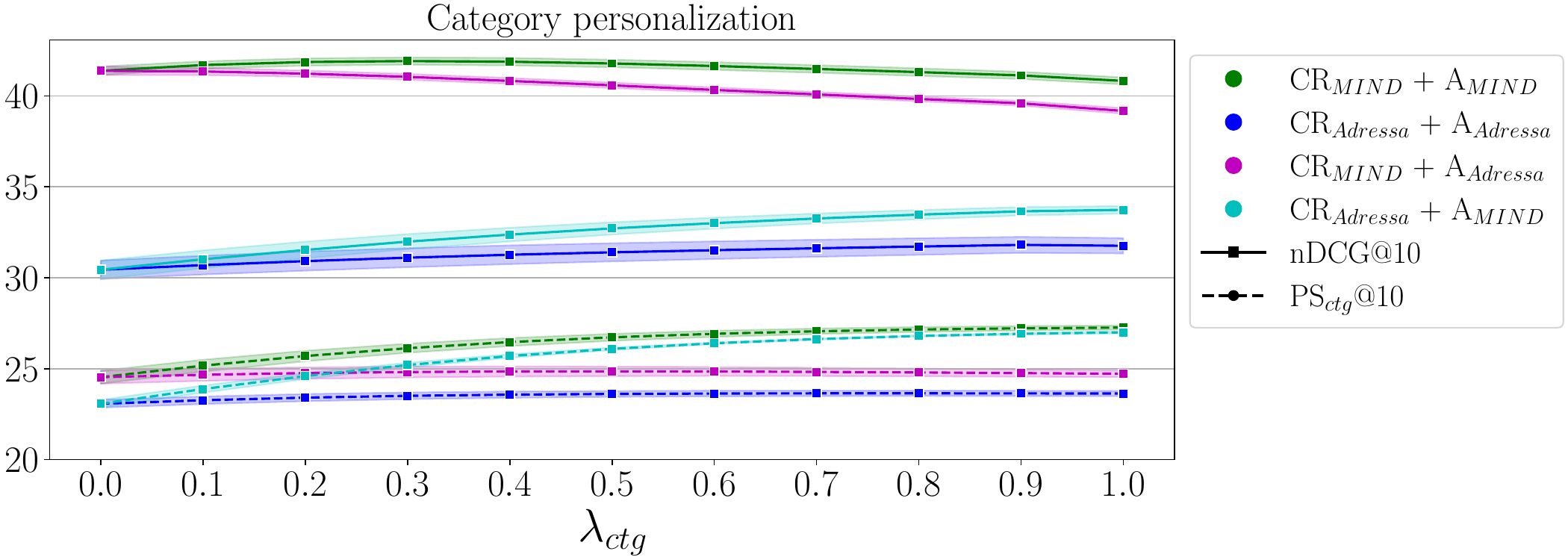}
         \caption{Category personalization.}
         \label{fig:adressa_transfer_mind_pers_categ}
     \end{subfigure}
    \caption{\texttt{XLT} in single-aspect personalization, with modules trained on different (combinations of) source-language datasets and evaluated on the target-language dataset MIND. The line style indicates the metric, the color the source-language datasets used in training. 
    }
    \label{fig:xlt_adressa_mind_pers}
    \vspace{-1em}
\end{figure*}

\begin{figure*}[h!]
     \centering
     \begin{subfigure}[b]{0.48\textwidth}
         \centering
         \includegraphics[width=\textwidth]{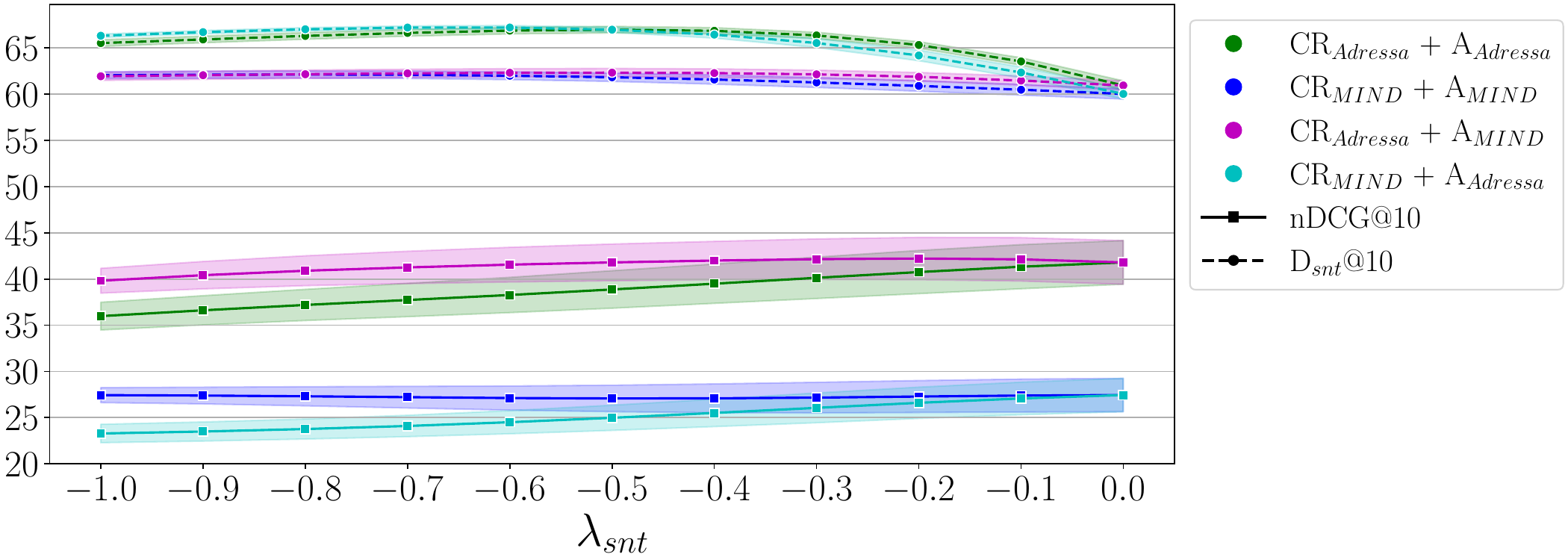}
         \caption{Sentiment diversification.}
         \label{fig:mind_transfer_adressa_div_sent}
     \end{subfigure}
     \hfill
     \begin{subfigure}[b]{0.48\textwidth}
         \centering
         \includegraphics[width=\textwidth]{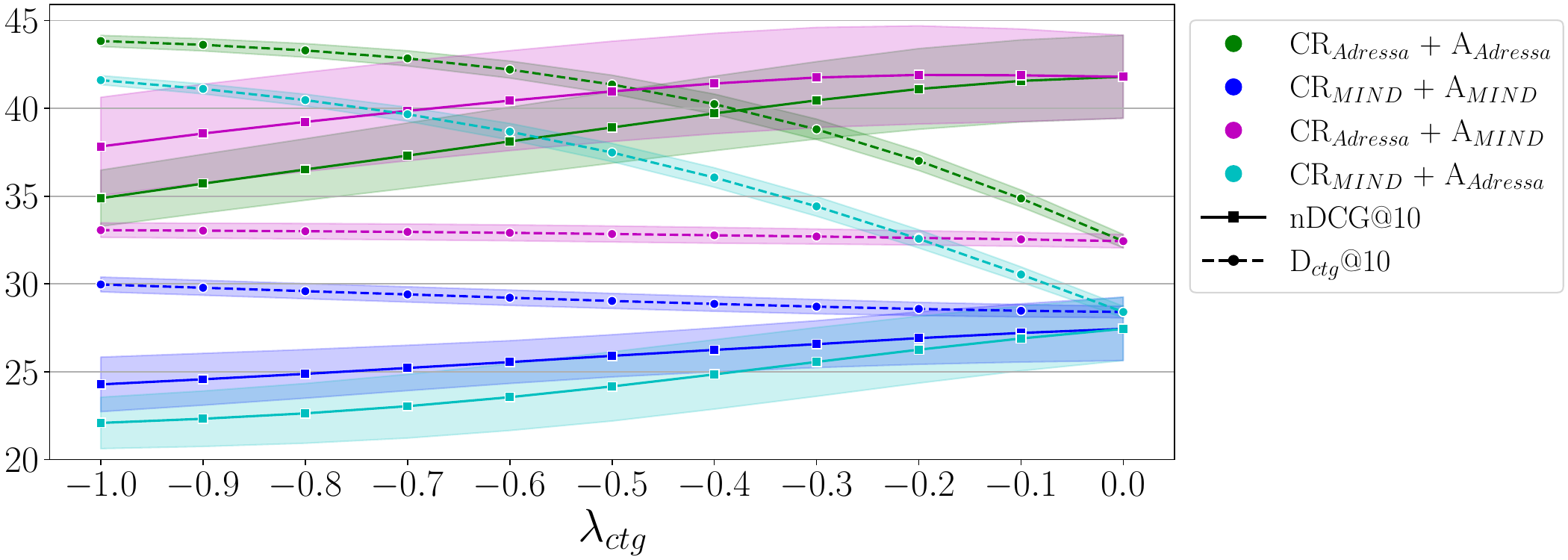}
         \caption{Category diversification.}
         \label{fig:mind_transfer_adressa_div_categ}
     \end{subfigure}
     \hfill
     \begin{subfigure}[b]{0.48\textwidth}
         \centering
         \includegraphics[width=\textwidth]{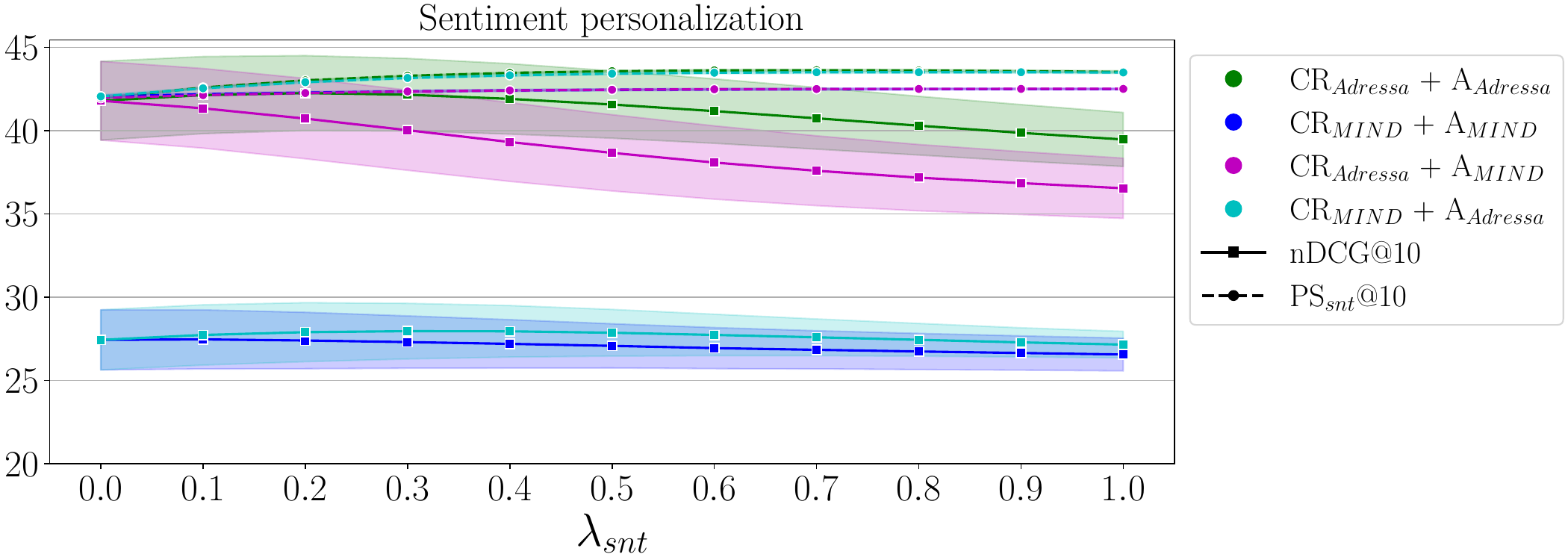}
         \caption{Sentiment personalization.}
         \label{fig:mind_transfer_adressa_pers_sent}
     \end{subfigure}
     \hfill
     \begin{subfigure}[b]{0.48\textwidth}
         \centering
         \includegraphics[width=\textwidth]{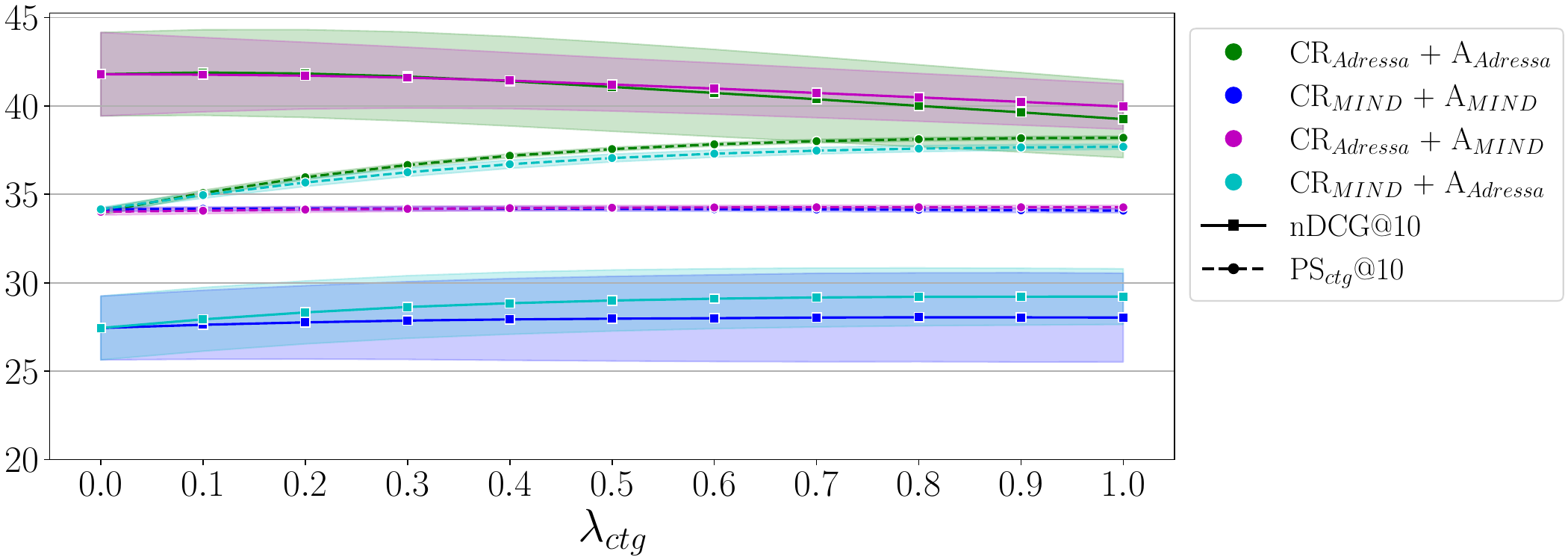}
         \caption{Category personalization.}
         \label{fig:mind_transfer_adressa_pers_categ}
     \end{subfigure}
    \caption{\texttt{XLT} in single-aspect diversification and personalization, with modules trained on different (combinations of) source-language datasets and evaluated on the target-language dataset Adressa. The line style indicates the metric, the color the source-language datasets used in training. 
    }
    \label{fig:xlt_mind_adressa}
    \vspace{-1em}
\end{figure*}

Fig. \ref{fig:xlt_adressa_mind_pers} summarizes the \texttt{XLT} results for single-aspect personalization on the target-language dataset MIND, whereas Fig. \ref{fig:xlt_mind_adressa} shows the analogous \texttt{XLT} results for single-aspect diversification and personalization, respectively, on the target-language dataset Adressa.

\subsection{Time Complexity Analysis}
\label{sec:appendix_time}

Table \ref{tab:inference_time} shows the average inference time for the entire MIND (365,201 impressions), and respectively, Adressa (146,284 impressions) test sets. Note that runtimes heavily depend on the computing infrastructure used, as well as parallel usage of the infrastructure for other tasks, as experiments are conducted on a HPC cluster. We highlight that \manner{} achieves a much lower inference time than the other NNRs.

\newcolumntype{g}{>{\columncolor{Gray}}r}
\begin{table}[h!]
\resizebox{\columnwidth}{!}{%
  \begin{tabular}{lll}
    \toprule
     \textbf{Model} & \textbf{MIND} & \textbf{Adressa}\\ \midrule
     NRMS-PLM & 17.53\textsubscript{$\pm$0.48}  &  7.13\textsubscript{$\pm$0.27}\\
     MINER & 16.03\textsubscript{$\pm$1.66}  & 9.96\textsubscript{$\pm$0.73} \\
     NAML-PLM & 33.99\textsubscript{$\pm$0.51}  & 7.09\textsubscript{$\pm$0.14} \\
     MINS-PLM & 27.50\textsubscript{$\pm$10.87}  & 7.81\textsubscript{$\pm$0.21} \\
     CAUM\textsubscript{no entities}-PLM & 22.67\textsubscript{$\pm$2.46}  & 8.12\textsubscript{$\pm$0.13} \\
     CAUM-PLM & 25.22\textsubscript{$\pm$0.45} & -- \\
     TANR-PLM & 17.02\textsubscript{$\pm$1.07}  & 6.98\textsubscript{$\pm$0.08} \\
     SentiRec-PLM &  17.93\textsubscript{$\pm$0.34} & 7.02\textsubscript{$\pm$0.08} \\
     SentiDebias-PLM & 21.01\textsubscript{$\pm$3.03}  & 13.28\textsubscript{$\pm$0.83} \\
     \manner{} (CR-Module) & 1.34\textsubscript{$\pm$0.03}  & 2.09\textsubscript{$\pm$0.06} \\
     \manner{} (CR-Module + \textit{ctg} A-Module) &  1.68\textsubscript{$\pm$0.08} & 2.78\textsubscript{$\pm$0.10} \\
     \manner{} (CR-Module + \textit{snt} A-Module) & 1.65\textsubscript{$\pm$0.01}  & 2.73\textsubscript{$\pm$0.06}\\
     \manner{} (CR-Module + 2 A-Modules) & 2.13\textsubscript{$\pm$0.05}  & 3.17\textsubscript{$\pm$0.01}\\
    
  \bottomrule
\end{tabular}%
}
\caption{Inference time (in thousands of seconds) for the different NNRs on the test portions of the MIND and Adressa datasets, respectively.}
\label{tab:inference_time}
\vspace{-0.5em}
\end{table}

\end{document}